\documentclass[preprint,showpacs,preprintnumbers,amsmath,amssymb,showkeys]{revtex4}

\usepackage[unicode,bookmarks,bookmarksopen,bookmarksopenlevel=0,colorlinks,%
pageanchor=1,breaklinks=1,linkcolor=blue,citecolor=blue,urlcolor=red,hypertex]{hyperref}

\usepackage{graphicx}
\usepackage{dcolumn}
\usepackage{bm}

\usepackage{float}

\begin{document}

\preprint{APS/123-QED}

\title{Mathematical Model of a pH-gradient Creation at~Isoelectrofocusing. \\ Part III: Numerical Solution of the Non-stationary Problem.}


\author{ E.\,V.~Shiryaeva}
\email{shir@math.sfedu.ru}

\author{N.\,M.~Zhukova}
\email{zhuk_nata@mail.ru}

\author{M.\,Yu.~Zhukov}%
 \email{myuzhukov@gmail.com}
\affiliation{%
Southern Federal University\\ Rostov-on-Don, Russia
}%

\date{\today}

\begin{abstract}
The mathematical model describing the non-stationary natural $\textrm{pH}$-gradient arising under the action of an electric field in an aqueous solution of ampholytes (amino acids)  is constructed and investigated. The model is part of a more general model of the isoelectrofocusing process. To numerical study of the model we use the finite elements method that allows to significantly reduce the computation time. We also examine the influence of the different effects (taking into account the water ions, the various forms of Ohm's law, the difference between isoelectric and isoionic points of the substances) on the process IEF.

\end{abstract}


\maketitle

\section{Introduction}\label{ZhS-01}

This paper continues the series of papers \cite{Part1,Part2} about $\textrm{pH}$-gradient creation at isoelectrofocusing (IEF). With the help of numerical methods we investigate non-stationary IEF problem in spatially one-dimensional case.
For the calculations the finite elements method (FEM) is applied. It allows to essentially reduce the time of calculations in comparison with widely used finite-difference methods (see, for example, \cite{Thormann2004,Thormann2006}). Moreover, the finite element method can be used for the solving two-dimensional and three-dimensional problems without significant modifications. The studying of the IEF by the numerical methods is not particularly new. However, in the numerical solution of the problem, as a rule, is limited by solving problems for certain concrete mixtures. It is still unclear how the mathematical model of the IEF process should be constructed accurate.
In particular, it is unclear in what form Ohm's law should be chosen. Whether to consider the diffusion electric current or it is enough to use only the electric current conduction. There is no response to the question about how important the presence of the water ions in a mixture.

Influence of the differences between isoelectric and isoionic points of the substances on the process IEF is not absolutely studied.
Recall that amphoteric substance actually have two characteristics: 1) the isoelectric points $\textrm{pI}$ at which the electrophoretic mobility of a substance is equal to zero; 2) the isoionic  points $\textrm{pI}_0$ at which the molar charge of a substance is equal to zero. The difference between these points is caused by the difference between mobilities of the negative and positive ions of the substance. Theoretical explanation of this fact is described in \cite{ZhukovBabskiyYudovich,BabZhukYudE}, and is experimentally confirmed, for example, in the T. Hirokawa tabels (see, for example, \cite{Hirokawa}. The difference ($\textrm{pI}$-$\textrm{pI}_0$) is very small.  It is considered that such difference should not significantly influence on the IEF. However, the obtained results show that it is not true. Indeed, in the initial stages of the process of focusing this effect is very weak. However, the difference ($\textrm{pI}$-$\textrm{pI}_0$) play important role when IEF process tends to stationary phase. In this case it is possible distortion the profiles of substances and $\textrm{pH}$-gradient.

The paper also refers to the importance of the use of precise numerical schemes for solving the electroneutrality equation. Insufficient accuracy of this equation solution leads to serious violations of some conservation laws, in particular, for the Kholraush's function.

The paper is organized as follows. In Sec.~\ref{App:ZhS-4} the basic equations of the non-stationary IEF problem are given.
In Sec.~\ref{App:ZhS-4.1} the numerical scheme of FEM is described. In Sec.~\ref{App:ZhS-4.2} the results of calculation are presented for five-component mixture.
The general Ohm's law is discussed in Sec.~\ref{App:ZhS-3}. The influence of the water ions on IEF is investigated in Sec.~\ref{App:ZhS-4.5}.
The influence of the differences between isoelectric and isoionic points on IEF is  discussed in Sec.~\ref{App:ZhS-4.6}.
Finally, in Sec.~\ref{App:ZhS-4.7} the implicit time discretization of the problem is demonstrated.

\section{Basic equations of the non-stationary IEF problem}\label{App:ZhS-4}

We restrict the study of the non-stationary IEF problem by 1D case. In dimensionless variables the equations system has the following form (see \cite{Part1,Part4,ZhukovBabskiyYudovich,BabZhukYudE,MosherSavilleThorman,Thormann2004,Thormann2006,Zhukov2005}):
\begin{equation}\label{ZhSeq-D1}
\partial_t a_{k}+\partial_x i_k=0,
\quad k=1,\dots,n,
\end{equation}
\begin{equation*}
i_k = -\varepsilon\mu_{k} \partial_x a_k+\mu_{k} \theta_{k}(\psi)a_k E,
\quad k=1,\dots,n,
\end{equation*}
\begin{equation}\label{ZhSeq-D2}
   \sum_{k=1}^{n}\theta_{k}(\psi)a_{k}=0,
\end{equation}
\begin{equation}\label{ZhSeq-D3}
j=\sum_{k=1}^{n}
\left(
-\varepsilon \mu_k \partial_x(\theta_{k}(\psi)a_k)+\mu_{k}\sigma_k(\psi)a_k E
\right),
\end{equation}
\begin{equation*}
 \partial_x j=0, \quad
E=-\partial_x \varphi,
\end{equation*}
\begin{equation}\label{ZhSeq-D4}
   \theta_{k}(\psi)= \frac{\sinh(\psi-\psi_{k})}{\cosh(\psi-\psi_{k})+\delta_{k}}, \quad
\end{equation}
\begin{equation*}
   \sigma_{k}(\psi)=\frac{\cosh(\psi-\psi_{k})}{\cosh(\psi-\psi_{k})+\delta_{k}}.
\end{equation*}
For constant voltage regime we take the boundary conditions corresponding to impermeability boundaries
\begin{equation}\label{ZhSeq-D5}
i_k\left|_{x=0}\right.=i_k\left|_{x=1}\right.=0, \quad k=1,\dots,n, \quad
\varphi\left|_{x=0}\right.=\varphi_0 \quad \varphi\left|_{x=1}\right.=0.
\end{equation}
The initial conditions has the form
\begin{equation}\label{ZhSeq-D6}
a_k\left|_{t=0}\right.=M_k(x), \quad k=1,\dots,n.
\end{equation}
Here, $a_k$, $i_k$ are the analytical concentration and the flux density of the components, $E$ is the intensity of external electric field, $j$ is the density of the electric current, $\psi$ is the acidity function of the mixture, $\mu_{k} \theta_{k}(\psi)$, $\mu_{k}\sigma_{k}(\psi)$, $\mu_{k}>0$, $\varepsilon \mu_k$ are the electrophoretic mobility, partial conductivity, characteristic mobility and diffusion coefficient of the components, $\varphi$ is the electric potential, $\varphi_0$ is the given potential difference, $M_k$ is the distribution of the concentration  $a_k$ quantity on the interval $[0,L]$, $\delta_k>0$ is the dimensionless parameter that characterize component, $\psi_k$ is the isoelectric point (electrophoretic mobility is equal to zero at $\psi=\psi_k$).

For convenience, we specify connection between dimensional and dimensionless variables (see also \cite{Part1}):
\begin {equation*}
  \widetilde{x} = xL_*, \quad
  \widetilde{t} = tt_*, \quad
  \widetilde{a} _k = a_k C_*, \quad
  \widetilde{E} = E E_*, \quad
  \widetilde{j} = j F_* C_* E_* \mu_*,
\end {equation*}
\begin {equation*}
  \varepsilon = \frac {R_*T_*}{F_* E_* L_*}, \quad
   {t_*} = \frac {L_*} {E_*\mu_*}.
\end {equation*}
Here, $L_*$, $t_*$, $E_*$, $C_*$ are the characteristic length, time, intensity of the electric field and analytical concentration; $\mu_*$ is the characteristic mobility; $F_*$ is the Faraday's number, $R_*$ is the universal gas constant, $T_*$ is the absolute temperature of the mixture.

\subsection{Parameters}\label{App:ZhS-4.1n}

The values of the parameters which we use for numerical calculation are presented in Tabs.~\ref{tab:table3}, \ref{tab:table4}  (see \cite{Righetti83}).
We chose these parameters the same as in \cite{Part2} to be able to compare the results of calculations for stationary and non-stationary problems.

\begin{table}[H]
\caption{\label{tab:table3} Dimensional parameters}
\begin{ruledtabular}
\begin{tabular}{lll}
$L_*$ & $0.1\,\textrm{m}$\quad ($0.0254\,\textrm{m}$) & Length \\
\hline
$C_*$ & $100\,\,\textrm{mol}/\textrm{m}^3=0.1\,\,\textrm{mol}/\textrm{litr} $ & Concentration \\
\hline
$\mu_*$ & $10^{-8}\,\,\textrm{m}^2/(\textrm{V}\cdot\textrm{s})$ & Electrophoretic mobility \\
\hline
$F_*$ & $96485.34\,\,\textrm{C}/\textrm{mol}$ & Faraday's number \\
\hline
$R_*$ & $8.314462\,\,\textrm{J}/(\textrm{mol}\cdot\textrm{K})$ & Universal gas constant \\
\hline
$T_*$ & $293\,\,\textrm{K}$ & Temperature \\
\hline
$\sigma_*=F_*C_*\mu_*$ & $0.09648534\,\textrm{C}/(\textrm{m}\cdot\textrm{V}\cdot\textrm{s})$ & Conductivity \\
\hline
$S_*$ & $ 10^{-5}\,\,\textrm{m}^2$ & Sectional area \\
\hline
$r_*$ & $ 0.0018\,\,\textrm{m}$ & Radius of capillary \\
\end{tabular}
\end{ruledtabular}
\end{table}

\begin{table}[H]
\caption{\label{tab:table4} Parameters of amino acids for five-component mixture (see \cite{Righetti83})}
\begin{ruledtabular}
\begin{tabular}{rccccccc}
 & $\textrm{pKb}_i$ & $\textrm{pKa}_i$ & $\textrm{pI}_i$ & $\psi_i$ & $\delta_i$  & $\mu_i$& $ M_i$\\
\hline
His-His         \quad ($a_1$) & $6.80$ & $7.80$ & $7.300$ & $-0.691$ & $1.58$   & $1.49$ & $0.2$ \\
His-Gly         \quad ($a_2$) & $6.27$ & $8.57$ & $7.420$ & $-0.967$ & $7.06$   & $2.40$ & $0.2$ \\
His             \quad ($a_3$) & $6.00$ & $9.17$ & $7.585$ & $-1.347$ & $19.23$  & $2.85$ & $0.3$ \\
$\beta$-Ala-His \quad ($a_4$) & $6.83$ & $9.51$ & $8.170$ & $-2.694$ & $10.94$  & $2.30$ & $0.1$ \\
Tyr-Arg         \quad ($a_5$) & $7.55$ & $9.80$ & $8.675$ & $-3.857$ & $6.67$   & $1.58$ & $0.1$ \\
\end{tabular}
\end{ruledtabular}
\end{table}

\section{Transformation to the variation form}\label{App:ZhS-4.1}

To solve the non-stationary IEF problem (\ref{ZhSeq-D1})--(\ref{ZhSeq-D6}) the finite elements method (FEM) is applied (\cite{FreeFem,ZhukShirFreeFem,FreeFemDoc}).
We transform the original problem to the variation form and use the semi-implicit time approximation.
We also use the following notification
\begin{equation}\label{ZhSeq-D7}
a_k^{(m)}=a_k^{(m)}(x)=a_k(x,t_m), \quad \varphi^{(m)}=\varphi^{(m)}(x)=\varphi(x,t_m),
\end{equation}
\begin{equation*}
\psi^{(m)}=\psi^{(m)}(x)=\psi(x,t_m),
\end{equation*}
where $t=m\tau$, $\tau$ is time step.

Semi-implicit time approximation has the form
\begin{equation}\label{ZhSeq-D8}
\frac{a_k^{(m+1)}-a_k^{(m)}}{\tau}+\partial_x i_k^{(m+1)}=0,
\end{equation}
\begin{equation}\label{ZhSeq-D9}
i_k^{(m+1)} = -\varepsilon\mu_{k} \partial_x a_k^{(m+1)}-\mu_{k} \theta_{k}(\psi^{(m)})a_k^{(m+1)} \partial_x\varphi^{(m)},
\end{equation}
\begin{equation}\label{ZhSeq-D10}
   \sum_{k=1}^{n}\theta_{k}(\psi^{(m)})a_{k}^{(m)}=0,
\end{equation}
\begin{equation}\label{ZhSeq-D11}
\partial_x \sum_{k=1}^{n}
\left(
\varepsilon \mu_k \partial_x(\theta_{k}(\psi^{(m)})a_k^{(m)})+\mu_{k}\sigma_k(\psi^{(m)})a_k^{(m)} \partial_x\varphi^{(m)}
\right)=0.
\end{equation}

To solve the algebraic equation (\ref{ZhSeq-D10}) (or (\ref{ZhSeq-D2})) we transform it to the following form (compare with (8), (11) in \cite{Part2})
\begin{equation}\label{ZhSeq-D12}
\psi=\frac12\ln
\frac{\displaystyle\sum\limits_{i=1}^n \frac{a_i e^{\psi_i}}{\varphi_i(\psi)}}
     {\displaystyle\sum\limits_{i=1}^n \frac{a_i  e^{-\psi_i}}{\varphi_i(\psi)}}
     \equiv G(\psi;a_1,\dots,a_n), \quad
  \varphi_i(\psi)=\cosh(\psi-\psi_i)+\delta_i.
\end{equation}
Obviously, the derivative with respect to $\psi$ is
\begin{equation*}
G'_\psi(\psi;a_1,\dots,a_n)=\frac12
\left\{
-\frac{\displaystyle\sum\limits_{i=1}^n \frac{a_i e^{\psi_i}}{\varphi_i(\psi)}\theta_i(\psi)}
     {\displaystyle\sum\limits_{i=1}^n \frac{a_i  e^{\psi_i}}{\varphi_i(\psi)}}
+\frac{\displaystyle\sum\limits_{i=1}^n \frac{a_i e^{-\psi_i}}{\varphi_i(\psi)}\theta_i(\psi)}
     {\displaystyle\sum\limits_{i=1}^n \frac{a_i  e^{-\psi_i}}{\varphi_i(\psi)}}
\right\}.
\end{equation*}
Taking into account
\begin{equation*}
\theta_i(\psi)<1, \quad a_i  \geqslant 0, \quad \sum\limits_{i=1}^n a_i^2 > 0,
\end{equation*}
we get
\begin{equation}\label{ZhSeq-D13}
|G'_\psi(\psi;a_1,\dots,a_n)| <1.
\end{equation}
It means that for determination of the  function $\psi^{(m)}(x)$ we can use the iteration method
\begin{equation}\label{ZhSeq-D14}
\psi^{(m),s+1}=|G(\psi^{(m),s};a_1^{(m)},\dots,a_n^{(m)})| <1, \quad s=0,1,\dots, \quad \psi^{(m),0}=\psi^{(m-1)}.
\end{equation}

The variation form of the problem (\ref{ZhSeq-D8}), (\ref{ZhSeq-D9}), (\ref{ZhSeq-D11}), (\ref{ZhSeq-D5}) is
\begin{equation}\label{ZhSeq-D15}
\int\limits_{0}^{1}\frac{a_k^{(m+1)}-a_k^{(m)}}{\tau} v_k(x)\,dx
-\int\limits_{0}^{1} i_k^{(m+1)} \partial_x v_k(x)\,dx=0,
\end{equation}
\begin{equation}\label{ZhSeq-D16}
i_k^{(m+1)} = -\varepsilon\mu_{k} \partial_x a_k^{(m+1)}-\mu_{k} \theta_{k}(\psi^{(m)})a_k^{(m+1)} \partial_x\varphi^{(m)},
\end{equation}
\begin{equation}\label{ZhSeq-D17}
\int\limits_{0}^{1} \partial_x \Phi(x) \sum_{k=1}^{n}
\left(
\varepsilon \mu_k \partial_x(\theta_{k}(\psi^{(m)})a_k^{(m)})+\mu_{k}\sigma_k(\psi^{(m)})a_k^{(m)} \partial_x\varphi^{(m)}
\right) dx=0,
\end{equation}
\begin{equation*}
\Phi(0)=\Phi(1)=0,
\end{equation*}
where $v_k(x)$, $\Phi(x)$ are the test functions.

\section{Numerical experiments. Five-component mixture}\label{App:ZhS-4.2}

For the five-component mixture the problem (\ref{ZhSeq-D14})--(\ref{ZhSeq-D17}) is solved numerically with the help of  $\textrm{FreeFem++}$ solver \cite{FreeFem,ZhukShirFreeFem,FreeFemDoc}. The physicochemical parameters are given in Tabs.~\ref{tab:table3} and \ref{tab:table4}.
To allow comparison of calculation results the composition of the mixture is chosen the same as in \cite{Part2}.

To calculate the dynamic of the IEF process we choose $200$ piecewise quadratic finite elements on interval $[0,1]$ and time step $\tau=0.005$.

\begin{figure}[H]
\centering
\includegraphics[scale=0.85]{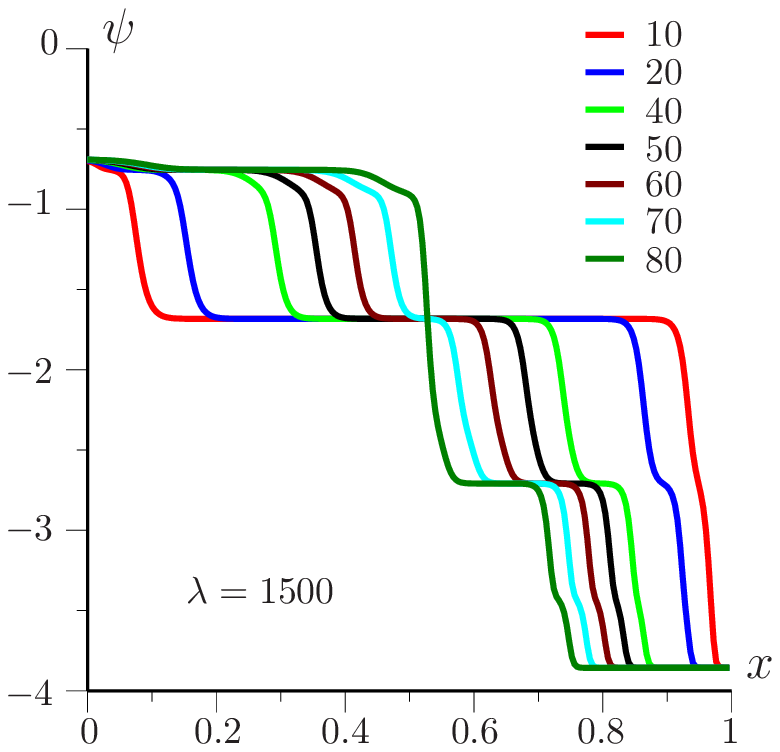}
\includegraphics[scale=0.85]{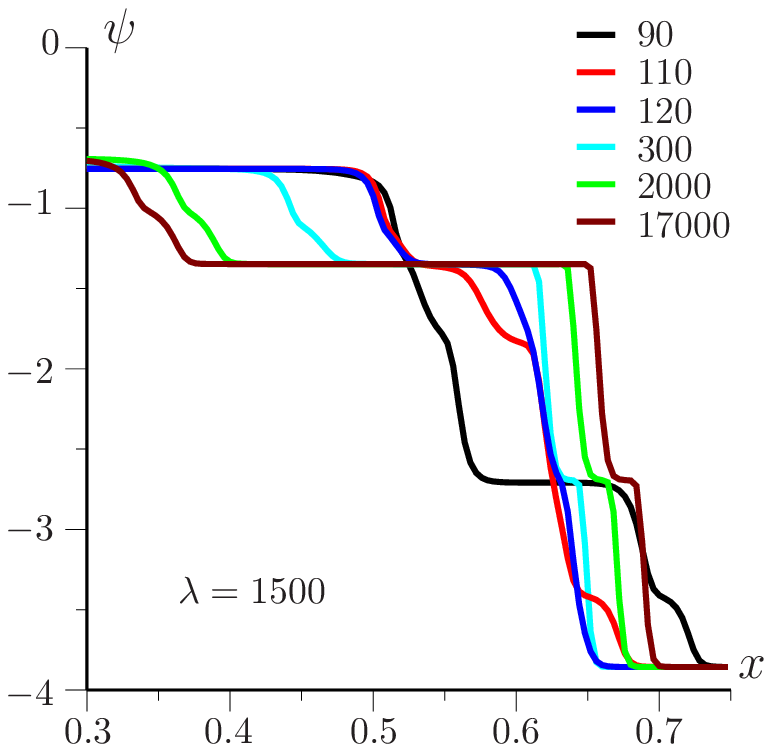}\\
\caption{Acidity $\psi(x)$ at $\lambda=1500$, $t=m\tau$, $\tau=0.005$}
\label{Ris18}
\end{figure}

On Fig.~\ref{Ris18} the results of $\textrm{pH}$-gradient calculation are presented at $\lambda=1500$ for
five-component mixture of the amino acids $\textrm{His-His}$, $\textrm{His-Gly}$, $\textrm{His}$, $\beta-\textrm{Ala-His}$, $\textrm{Tyr-Arg}$ with initial concentrations:
\begin{equation}\label{ZhSeq-D18}
M_1=0.5, \quad M_2=0.05, \quad M_3=0.5, \quad M_4=0.05, \quad M_5=0.5.
\end{equation}
We specify the parameters $\varphi_0$ as $U_0$ for stationary state at $\lambda=1500$ (see caption of the Fig.~5 in \cite{Part2})
\begin{equation}\label{ZhSeq-D19}
\varphi_0=U_0=2.808, \quad (j_0=1, \quad \varepsilon=1/\lambda).
\end{equation}
The characteristics time $t_*$, current density $j_*$, electric current $I_*$, and voltage $U_*$ are determined by the formulae
\begin{equation}\label{ZhSeq-D20}
U_*=\frac{R_*T_*}{F_*}\lambda \approx 0.025\,\frac{\lambda}{j_0}\,(V) , \quad t_*=\frac{L_*}{\mu_*U_*}\approx 3.96\cdot 10^7 \frac{j_0}{\lambda}\,(s),
\end{equation}
\begin{equation}\label{ZhSeq-B9}
  j_*=\lambda \frac{R_*T_*C_*\mu_*}{L_*}\approx 0.025\,\lambda\,\,(\textrm{A}\cdot\textrm{m}^{-2}), \quad
  I_*=j_*S_*\approx 0.25\cdot10^{-6}\,\lambda (\,\,\textrm{A}).
\end{equation}

In particular, at $\lambda=1500$ the dimension value of the $\tau$ is
\begin{equation}\label{ZhSeq-D21}
\tau_*=\tau t_*\approx 132\,(s).
\end{equation}

The Fig.~\ref{Ris18} shows clearly the process of a stepped-$\textrm{pH}$-gradient formation over time. The concentrations $a_2$ and $a_4$ are small and their influence on the $\textrm{pH}$-gradient is weak. Over the time interval  $t\in [0, 90\tau$, so-called cathode drift $\textrm{pH}$-gradient is observed \cite{Righetti83}.  Other words,
the  $\textrm{pH}$-gradient moves to right side ($x=1$). The most smooth profile of the $\textrm{pH}$-gradient is observed at $t=110\tau$ (red line on right Fig.~\ref{Ris18}). At next stages of the $\textrm{pH}$-gradient formation the concentrations $a_2$ and $a_4$ are playing significant role, forming on  $\textrm{pH}$-gradient profile additional step (starting from $t=300\tau$ at $x\approx 0.36$ and $x\approx 0.66$). The final stage of the process, when the gradient is close to stationary (compare with Fig.~5 in \cite{Part2}),  comes at the moment $t=17000\tau$ ($632$ hours approximately!). Of course, this time is difficult to implement in a real experiment. A significant decrease of the time of $\textrm{pH}$-gradient formation can be achieved by reducing the length of the electrophoretic chamber, for example, realizing the IEF process in a micro-device. For instance, in \cite{Santiago2003} the micro-device with length $2.54\,\,\textrm{cm}$, width $200\,\,\mu\textrm{m}$, and depth $20\,\,\mu\textrm{m}$ is described.

To avoid misunderstandings, it is appropriate to mention that the main objective of the research is the demonstration of the  mathematical models and methods efficiency. Although, the real parameters are used, the presented numerical examples demonstrate the ability of the stepped-$\textrm{pH}$-gradients calculation and explore dynamics of the IEF process. Unrealistically large time of the transition to a steady state process primarily indicates the need of the mathematical modeling of the IEF process. These calculations allow to predict the experiment and give interpretation of the obtained results. This is all the more important for  many of the physical characteristics of the process such as the conductivity of mixture, electric field strength, and $\textrm{pH}$-gradient, in practice it does not measurable inside the electrophoretic chamber. Usually, experimental measurements, which characterize the process, only register the location of the samples in the electrophoretic chamber.

\begin{figure}[H]
\centering
\includegraphics[scale=0.80]{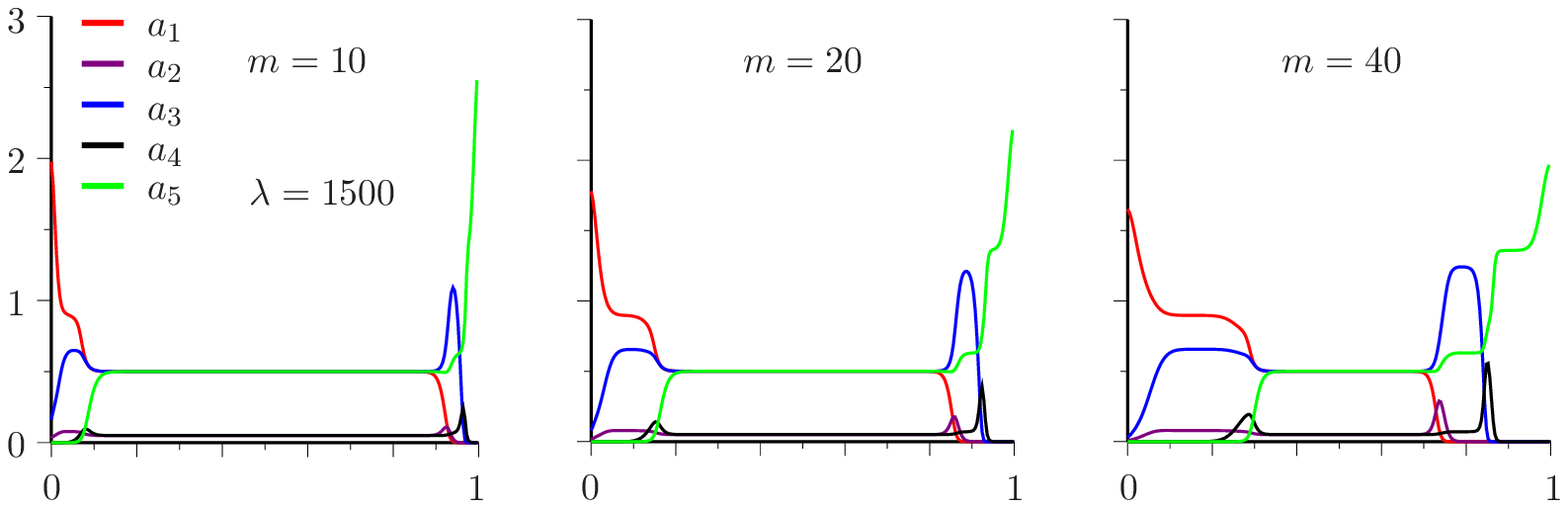}\\ [-4mm]
\includegraphics[scale=0.80]{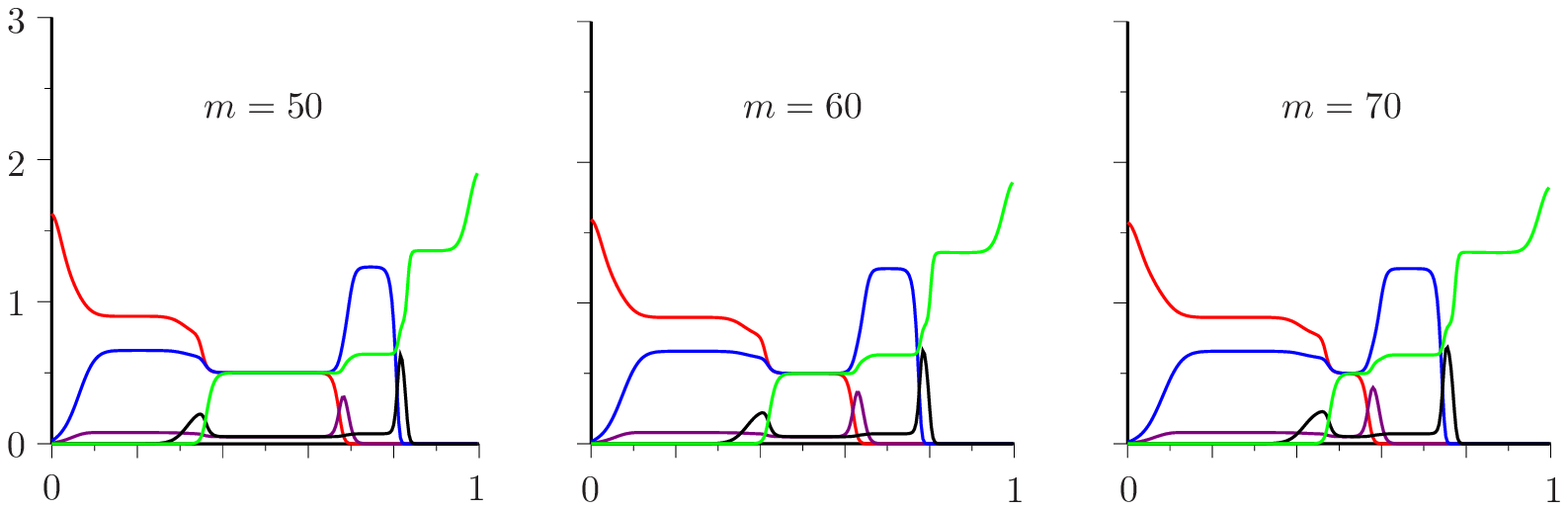}\\ [-4mm]
\includegraphics[scale=0.80]{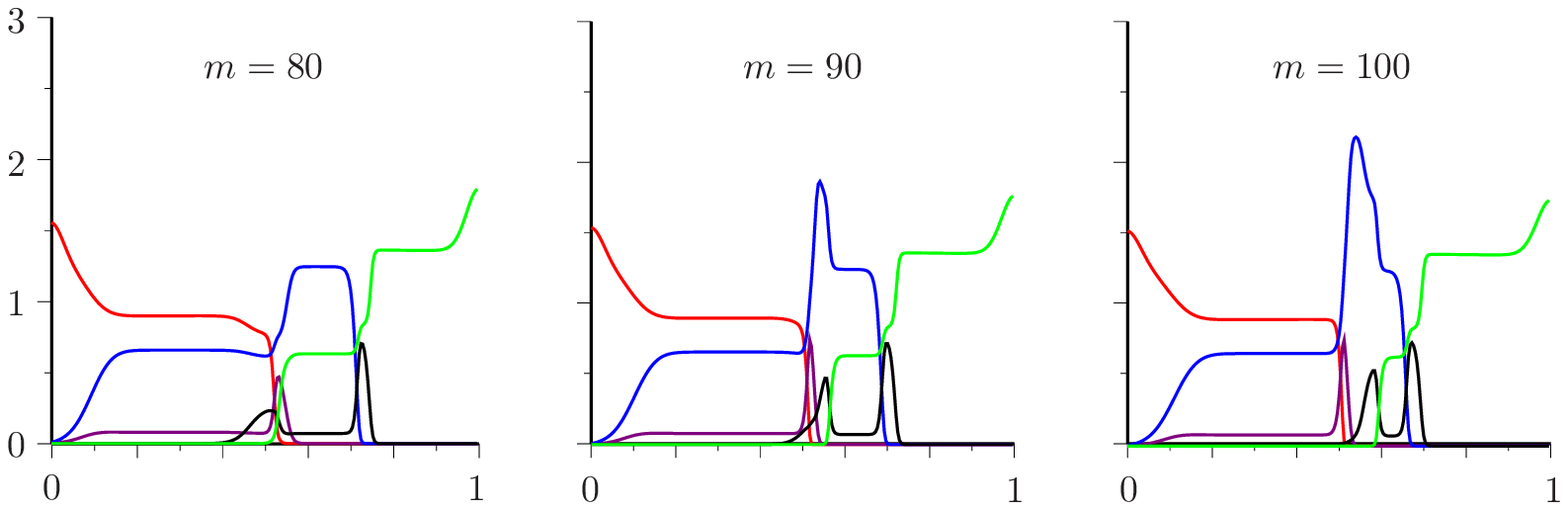}\\ [-4mm]
\includegraphics[scale=0.80]{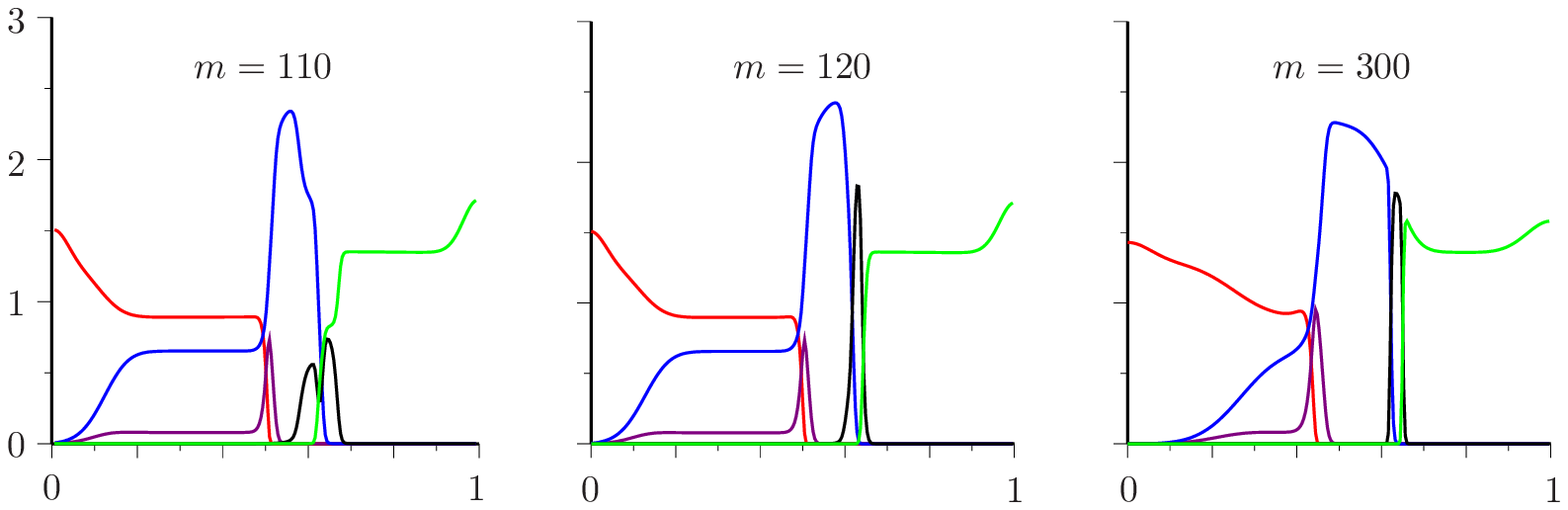}\\ [-4mm]
\includegraphics[scale=0.80]{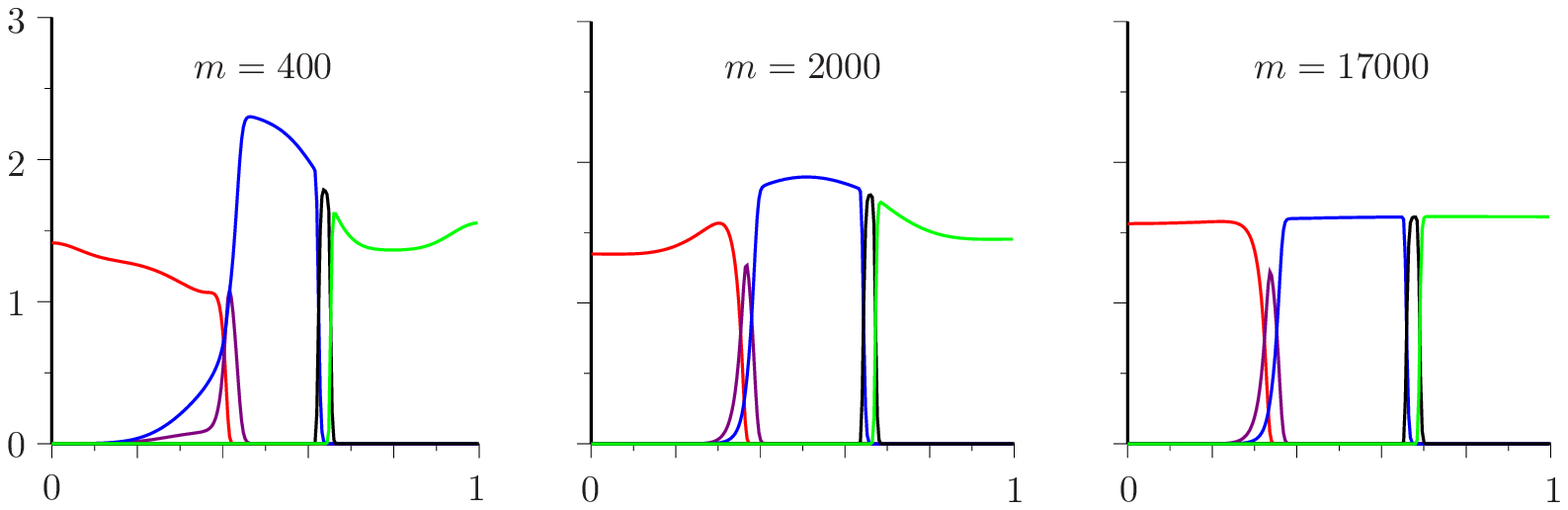}\\ [-6mm]
\caption{Concentration $a_k(x)$ at $\lambda=1500$, $t=m\tau$, $\tau=0.005$}
\label{Ris19}
\end{figure}

Fig.~\ref{Ris19} demonstrates the concentration evolution over time from $t=0$ to $t=17000\tau$ (stationary  state) at $\lambda=1500$.
On Fig.~\ref{Ris19} the distribution of the concentrations at different time moment is shown. The process of the $\textrm{pH}$-gradient creation can be interpreted as the fractionation of the mixture into individual components. After separation the each individual concentration corresponds to a sloping part of the $\textrm{pH}$-gradient. On the boundary between two concentrations the $\textrm{pH}$-gradient jump is observed. If IEF process is interpreted as fractionation of the mixture then one can
observe the three peaks starting from $t=70\tau$. The middle one corresponds to $a_2$ and two other peaks corresponds to $a_4$.

Experimenter can be expected by mistake
that these two peaks correspond two different substances. Three peaks instead of two are observed in the over time from $t=20\,\tau$ to $t=120\,\tau$.
The concentrations of substances $a_2$, $a_4$ are small compared with concentrations $a_1$, $a_3$, $a_5$. This means that the $\textrm{pH}$-gradient is created by substances $a_1$, $a_3$, and $a_5$. The substances $a_2$ and $a_4$ only slightly distort this $\textrm{pH}$-gradient. Stationary state is achieved at the time moment $t\approx 17000\,\tau$.

\begin{figure}[H]
\centering
\includegraphics[scale=0.9]{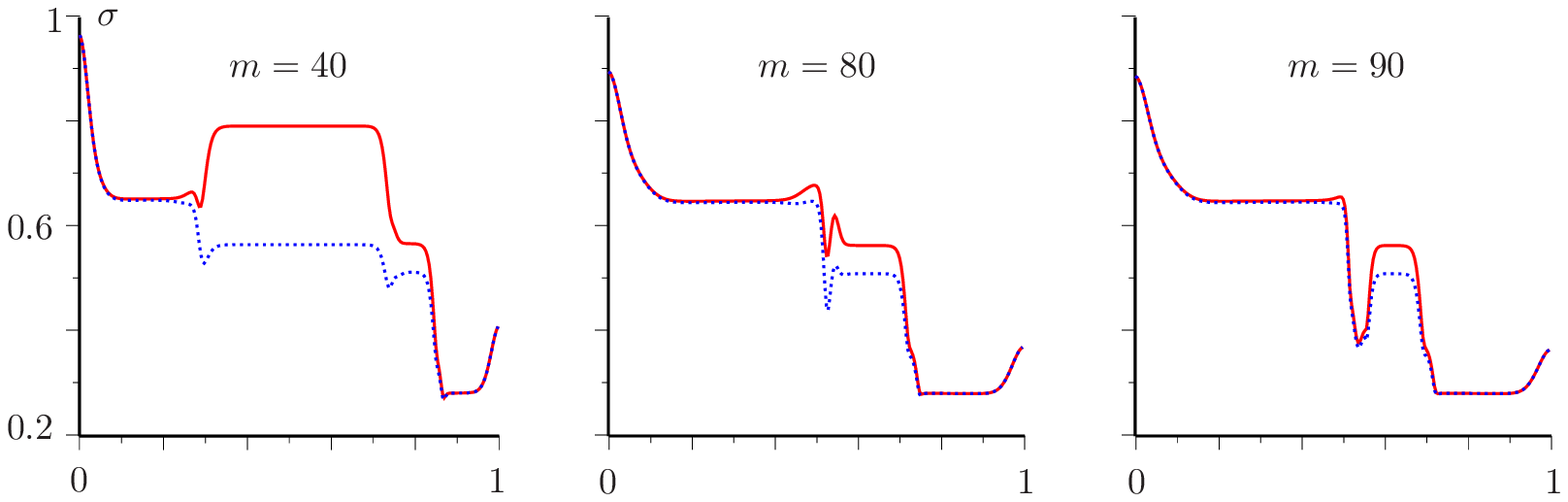}\\ [-4mm]
\includegraphics[scale=0.9]{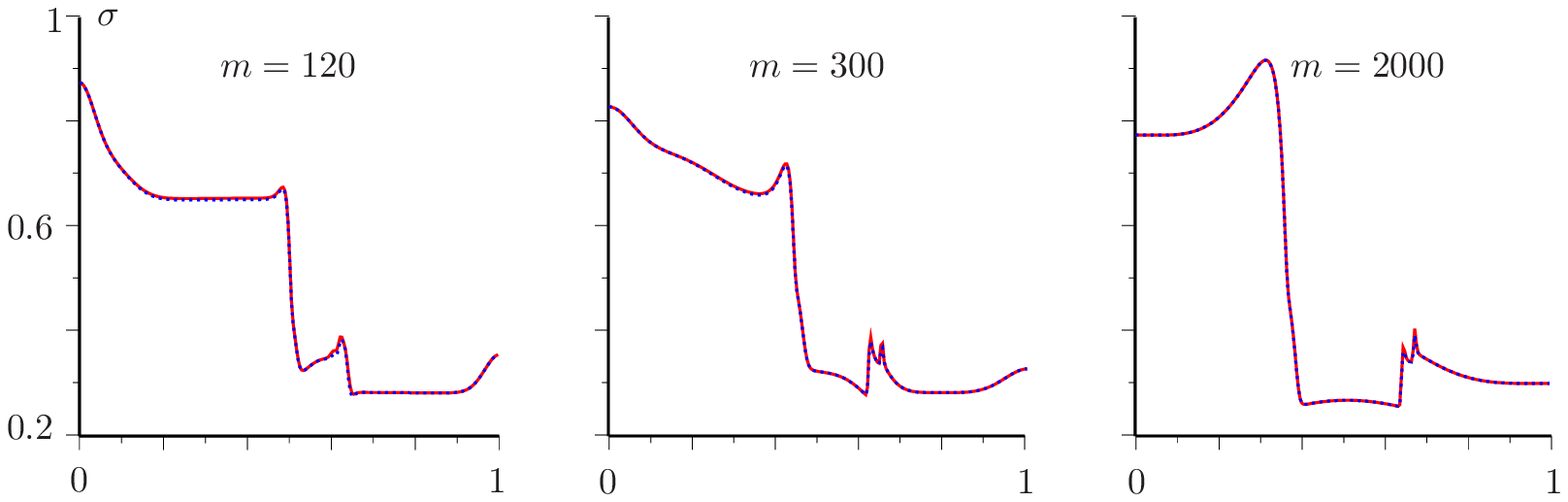}\\ [-4mm]
\caption{Conductivity $\sigma(x)$ at $\lambda=1500$, $t=m\tau$, $\tau=0.005$. Solid line corresponds to general Ohm's law. Dotted line corresponds to usual Ohm's law}
\label{Ris20}
\end{figure}

On the Fig.~\ref{Ris20} the distribution of the mixture conductivity across electrophoretic chamber is shown at different time moments. This figure demonstrates the difference between generalized and usual Ohm's law (see Appendix~\ref{App:ZhS-3}). Especially this effect is observed in time interval $t\in [40\tau,90\tau]$.
This fact proves the need to correct choice of the Ohm's law for modeling IEF process.

\begin{figure}[H]
\centering
\includegraphics[scale=0.9]{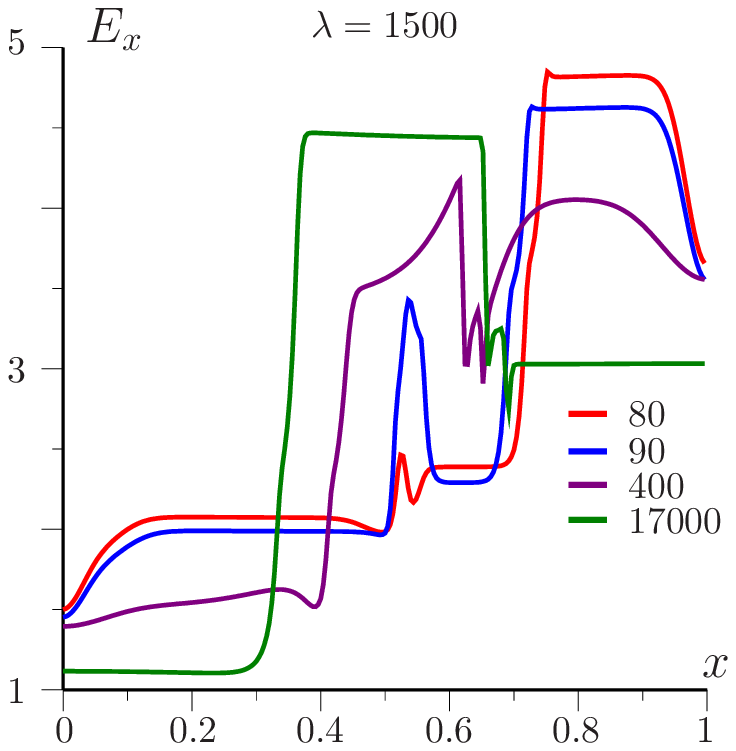}\
\includegraphics[scale=0.9]{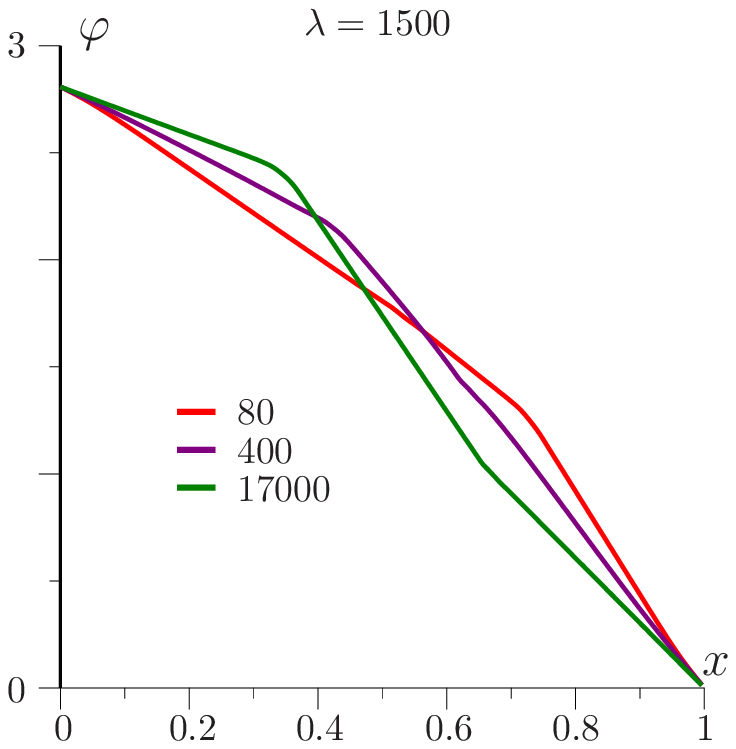}\\  [-6mm]
\caption{Intensity of electric field $E_x(x)$ and potential $\varphi(x)$ at $\lambda=1500$, $t=m\tau$, $\tau=0.005$}
\label{Ris21}
\end{figure}

On the Fig.~\ref{Ris21} the distribution of the electric field intensity and
electric potential are shown at different time moments. This figure demonstrates the
complicated behavior  of the electric field inside of the electrophoretic chamber.

\subsection{Numerical experiments. Samples injection}\label{App:ZhS-4.3}

In this section we present the numerical solution of all the problem for the following initial data:
\begin{equation}\label{ZhSeq-D22}
  a_1\bigl|_{t=0}=M_1, \quad
  a_2\bigl|_{t=0}=0, \quad
  a_3\bigl|_{t=0}=M_3, \quad
  a_4\bigl|_{t=0}=0, \quad
  a_5\bigl|_{t=0}=M_5.
\end{equation}
We solve the problem (\ref{ZhSeq-D1})--(\ref{ZhSeq-D5}), (\ref{ZhSeq-D22}) (or
(\ref{ZhSeq-D14})--(\ref{ZhSeq-D17}), (\ref{ZhSeq-D22})) on the time interval $[0,500]$.

After that we replace initial data (\ref{ZhSeq-D22}) by the following data:
\begin{equation}\label{ZhSeq-D23}
  a_1\bigl|_{t=t_0}=a_1\bigl|_{t=t_0-0}, \quad
  a_3\bigl|_{t=t_0}=a_3\bigl|_{t=t_0-0}, \quad
  a_5\bigl|_{t=t_0}=a_5\bigl|_{t=t_0-0}, \quad x\in [0,1],
\end{equation}
\begin{equation}\label{ZhSeq-D24}
  a_2\bigl|_{t=t_0}=\left\{
  \begin{array}{cl}
    M_2/x_0, & x\in [0,x_0],\\
    0, & x\in (x_0,1],
  \end{array}
  \right.\qquad
  a_4\bigl|_{t=t_0}=\left\{
  \begin{array}{cl}
    M_4/x_0, & x\in [0,x_0],\\
    0, & x\in (x_0,1],
  \end{array}
  \right\}
\end{equation}
where
\begin{equation*}
  t_0=500\,\tau,\quad
  t_0-0=499\,\tau,\quad
  x_0=0.1.
\end{equation*}

Formulae (\ref{ZhSeq-D24}) simulate the injection of the samples
toward the interior of the electrophoretic chamber. Non-stationary $\textrm{pH}$-gradient is created during $t=500\tau$ with the help of the concentration $a_1$, $a_3$, $a_5$. The separation of the samples $a_2$, $a_4$ occurs in the given non-stationary $\textrm{pH}$-gradient.
In this case, we do not observe the appearance of the three peaks as in the previous example (see Fig.~\ref{Ris19}).
Complete separation of the mixture comes at time $t=2000\,\tau$ approximately.
In this time the concentration are embedded in a previously created $\textrm{pH}$-gradient.

\begin{figure}[H]
\centering
\includegraphics[scale=0.85]{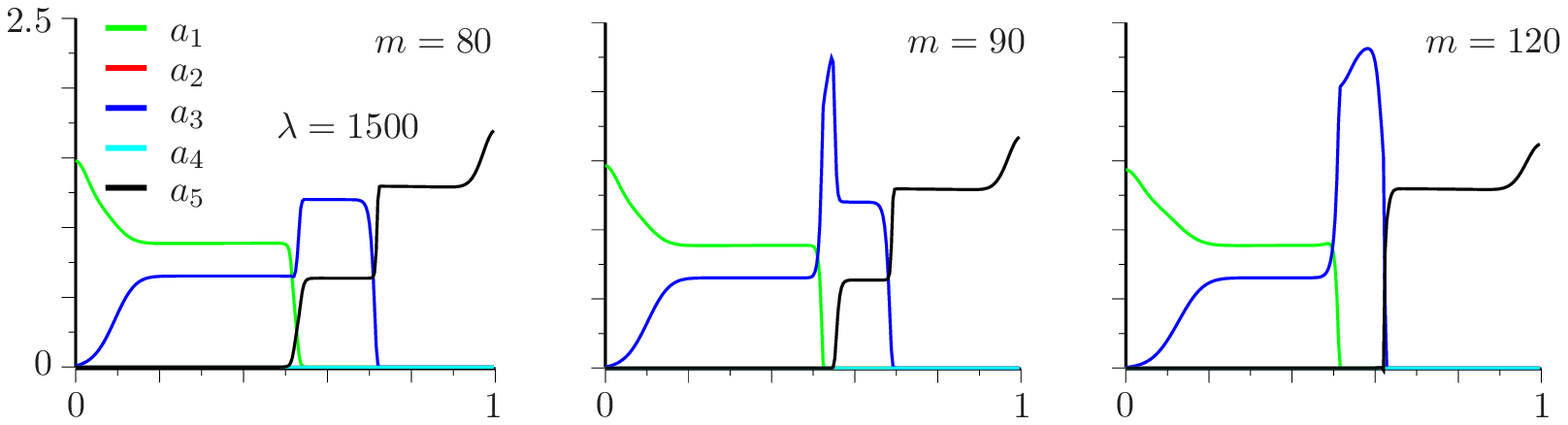}\\ [-4mm]
\includegraphics[scale=0.85]{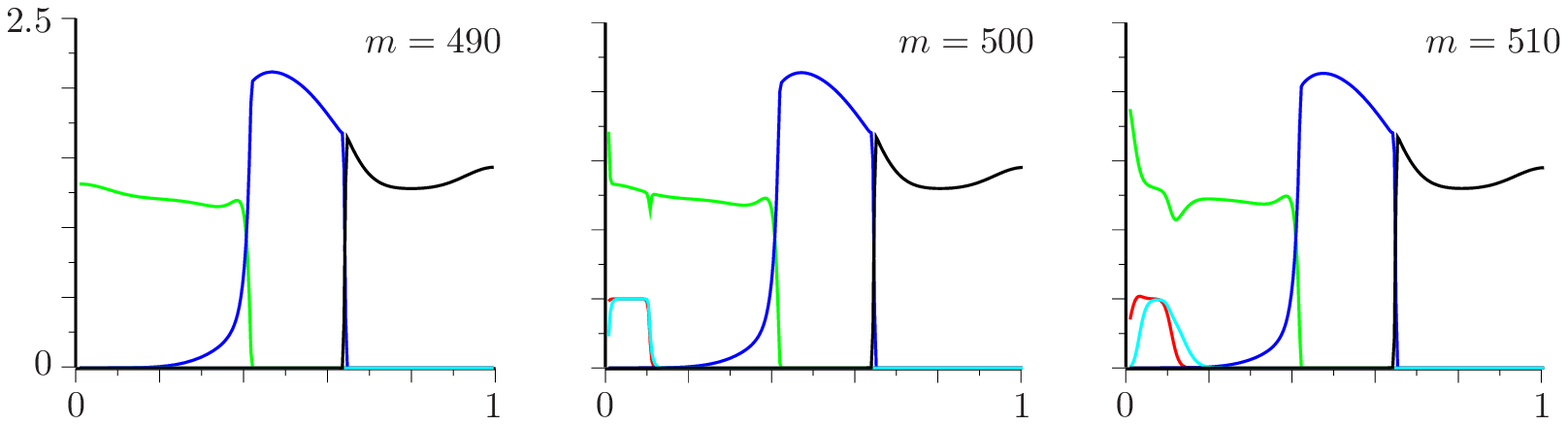}\\  [-4mm]
\includegraphics[scale=0.85]{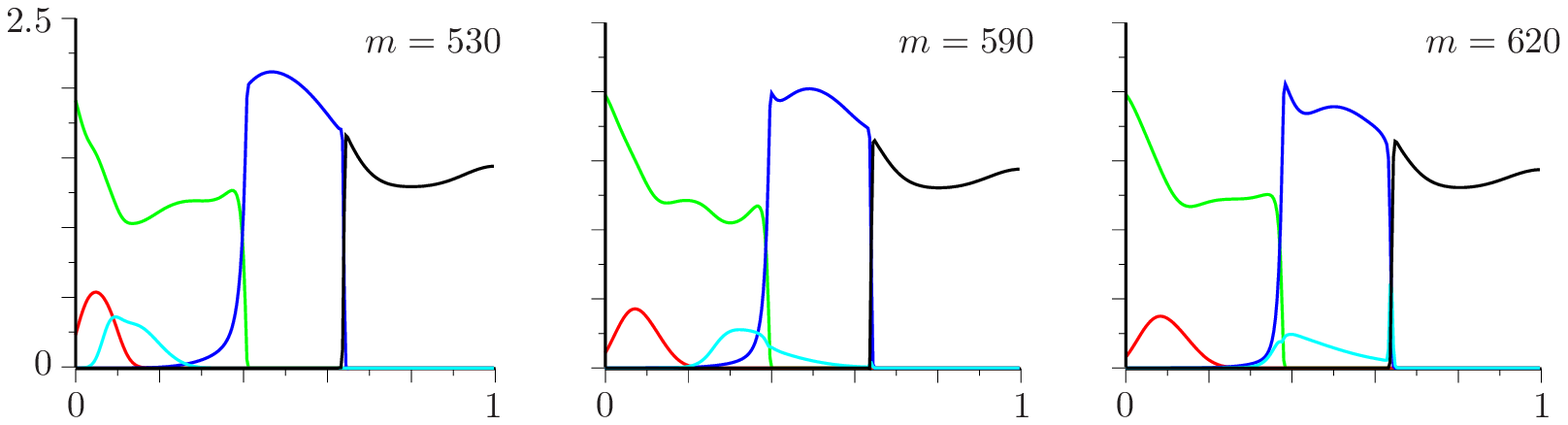}\\  [-4mm]
\includegraphics[scale=0.85]{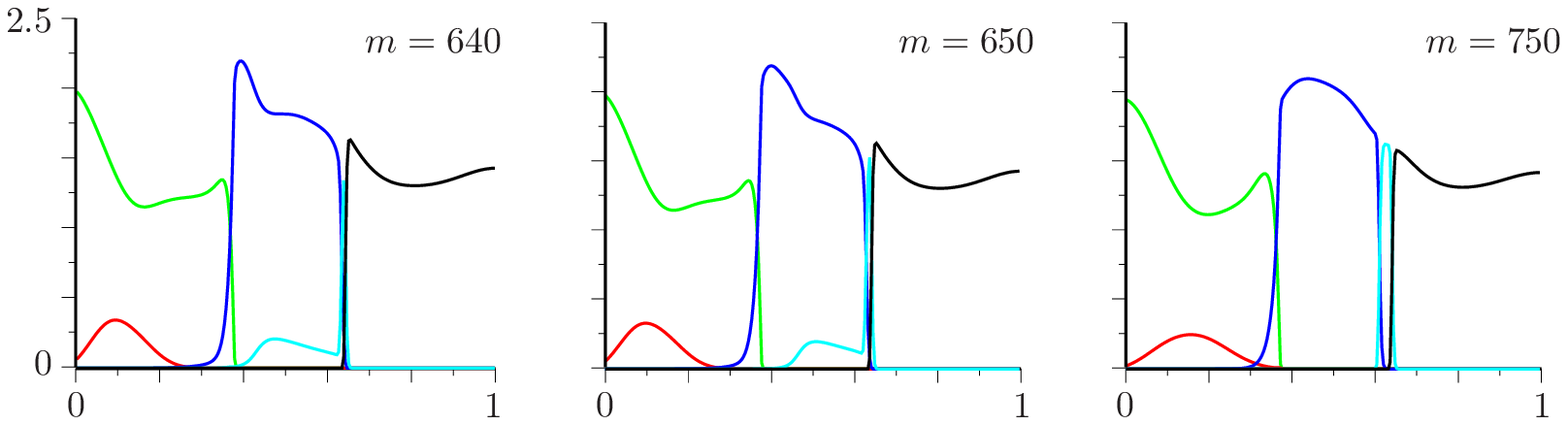}\\ [-4mm]
\includegraphics[scale=0.85]{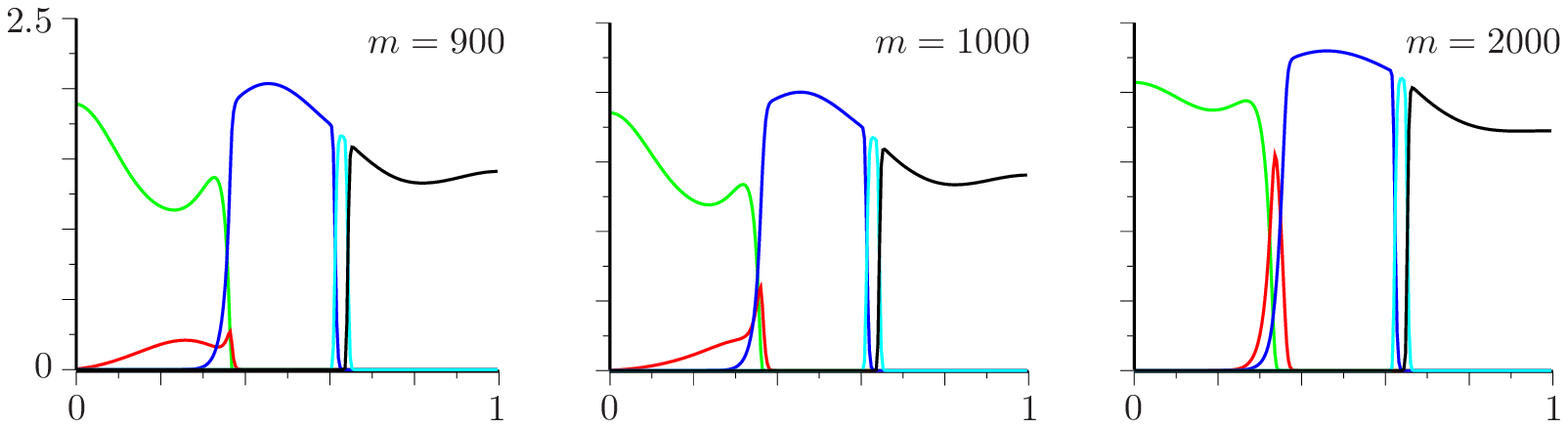}\\ [-6mm]
\caption{Concentration $a_k(x)$ at $\lambda=1500$, $t=m\tau$, $\tau=0.005$}
\label{Ris22} 
\end{figure}

\begin{figure}[H]
\centering
\includegraphics[scale=0.85]{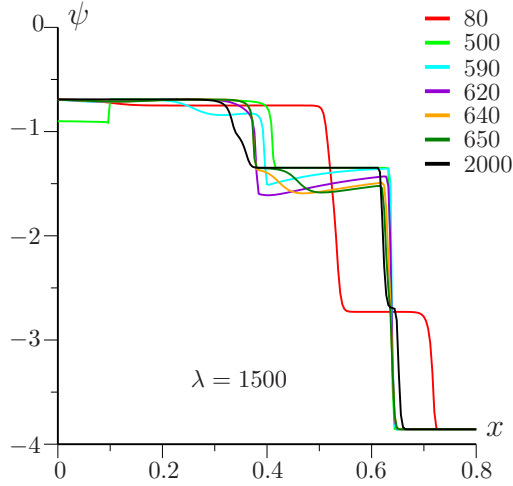}\\  [-4mm]
\caption{Acidity $\psi(x)$ at $\lambda=1500$, $t=m\tau$, $\tau=0.005$}
\label{Ris23}
\end{figure}

In the case of initial data (\ref{ZhSeq-D23}), (\ref{ZhSeq-D24})  the evolution of the $\textrm{pH}$-gradient is shown on Fig.~\ref{Ris23}.

\section{On generalized Ohm's law}\label{App:ZhS-3}

At construction of the mathematical model (\ref{ZhSeq-D1})--(\ref{ZhSeq-D6}) a generalized Ohm's law (\ref{ZhSeq-D3}) is used. We assume that the total current density $\boldsymbol{j}$ consists of the conduction current density  $\boldsymbol{j}_{\textrm{cond}}$ and electric diffusion current density $\boldsymbol{j}_{\textrm{diff}}$
\begin{equation}\label{ZhSeq-C1}
\boldsymbol{j}=\boldsymbol{j}_{\textrm{cond}}+\boldsymbol{j}_{\textrm{diff}},
\end{equation}
\begin{equation*}
\boldsymbol{j}_{\textrm{cond}}=
\sum_{k=1}^{n}
\mu_{k}\sigma_k(\psi)a_k \boldsymbol{E}, \quad
\boldsymbol{j}_{\textrm{diff}}=
-\sum_{k=1}^{n}
\varepsilon \mu_k\nabla(e_{k}(\psi)a_k).
\end{equation*}

In the case of the usual Ohm's law diffusion current $\boldsymbol{j}_{\textrm{diff}}$ is omitted. In this case the conductivity of the mixture is determined by the relation
\begin{equation}\label{ZhSeq-C2}
\sigma=
\sum_{k=1}^{n}
\mu_{k}\sigma_k(\psi)a_k.
\end{equation}
In the case of a generalized law and stationary problem we have the formula (see formula (10) in \cite{Part1})
\begin{equation}\label{ZhSeq-C3}
\sigma_{\textrm{stat}}=
\sum_{k=1}^{n}
\mu_{k}\theta'_k(\psi)a_k
\end{equation}
instead the formula (\ref{ZhSeq-C2}).

\begin{figure}[H]
\centering
\includegraphics[scale=0.85]{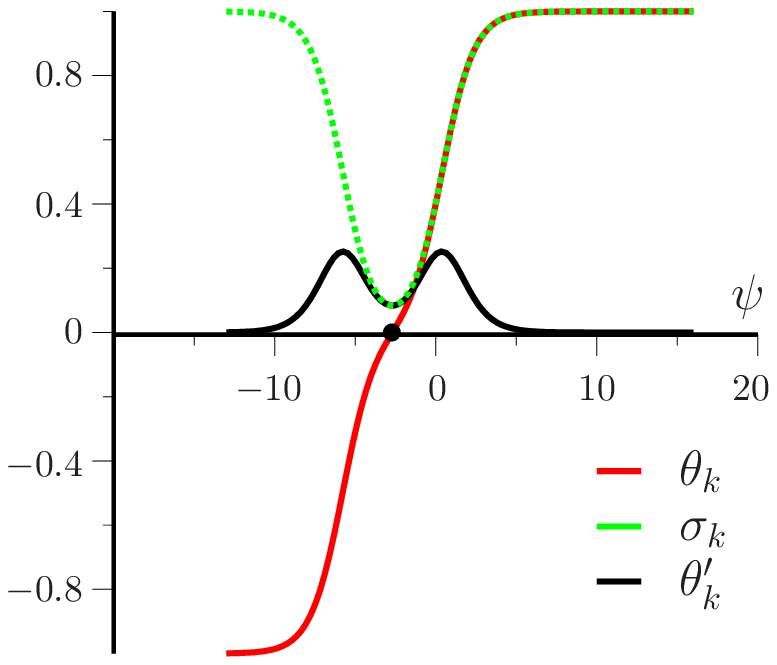}
\includegraphics[scale=0.85]{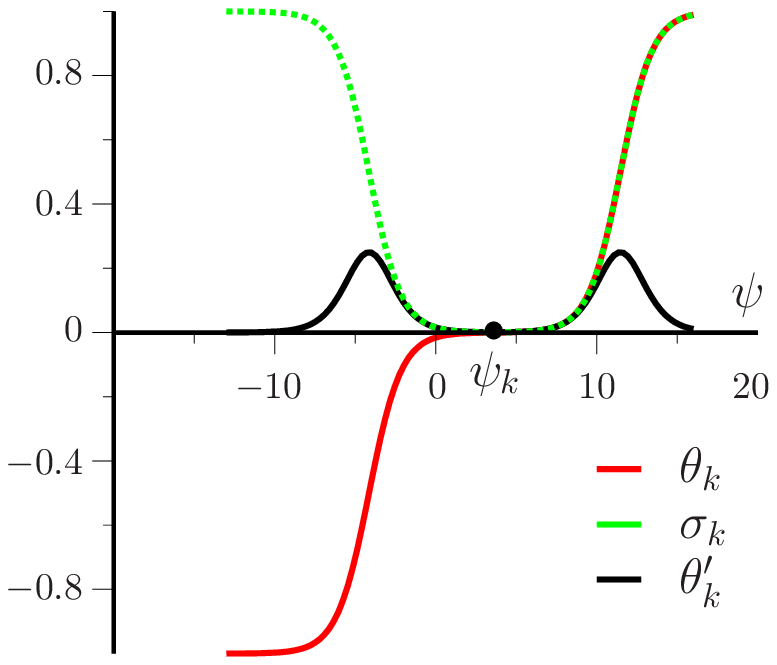}\\  [-4mm]
\caption{Dependences $\sigma_k$, $\theta_k$, $\theta'_k$ on $\psi$ for $\beta$-Ala-His (left) and
Asn (right), see table~\ref{tab:table4}}
\label{Ris15}
\end{figure}

The role of the specific conductivity plays $\mu_k\theta'_k(\psi)$ which is called effective specific conductivity. On Fig.~\ref{Ris15} the differences between $\mu_k\theta'_k(\psi)$ and $\mu_k\sigma_k(\psi)$ are shown.

In the vicinity of the isoelectric point $\psi_k$ the specific conductivity $\mu_k\sigma_k$ is small, but outside of this vicinity $\mu_k\sigma_k$  is almost constant. On the contrary, effective specific conductivity $\mu_k\theta'_k$ practically equals to zero outside of the isoelectric point vicinity.

\begin{figure}[H]
\centering
\includegraphics[scale=0.85]{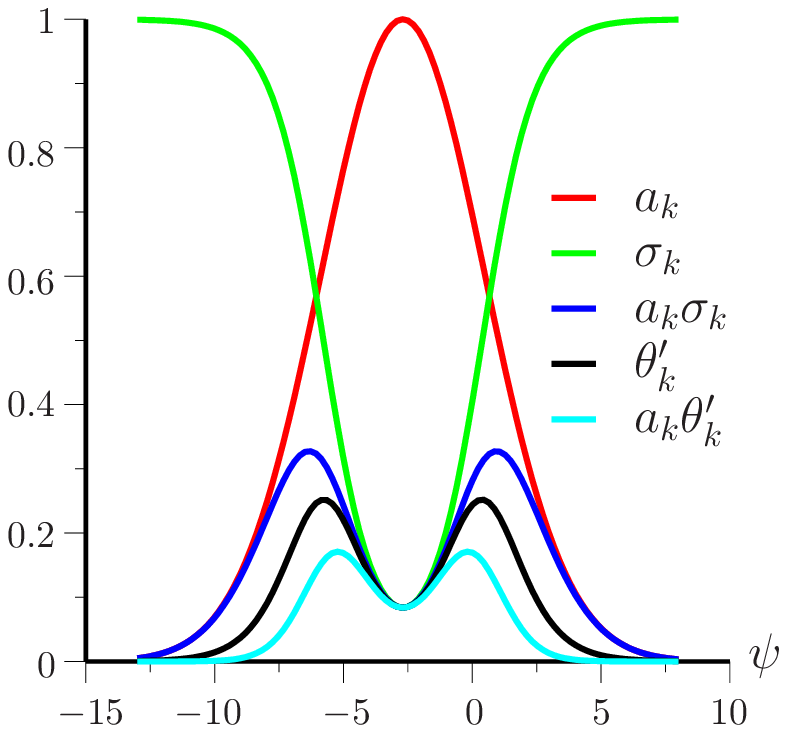}
\includegraphics[scale=0.85]{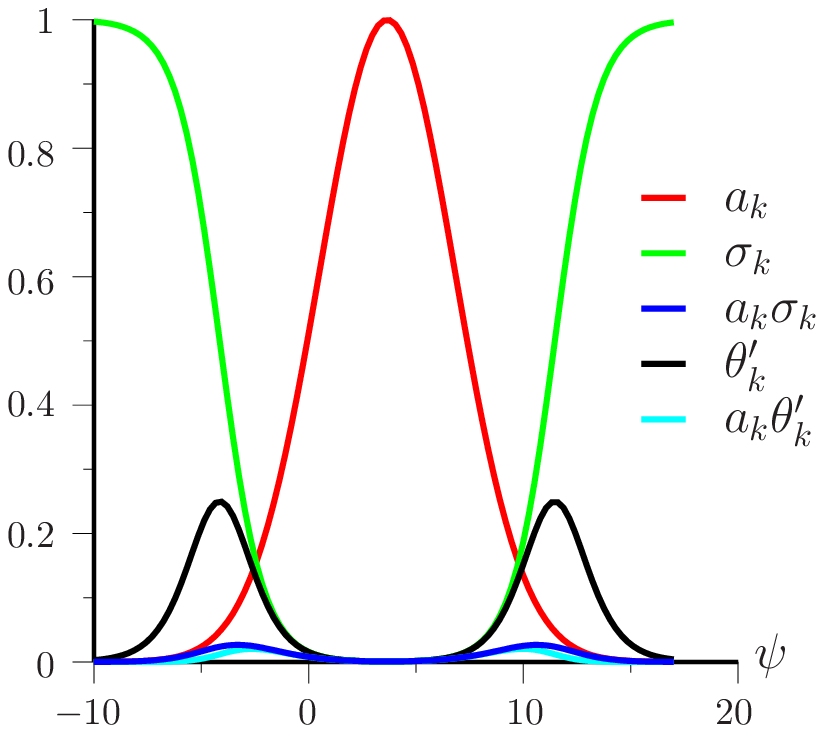}\\ [-4mm]
\caption{Dependence $a_k$, $\sigma_k$, $a_k\sigma_k$,
$\theta'_k$, $a_k\theta'_k$ on $\psi$ for $\beta$-Ala-His (left) and
Asn (right), see table~\ref{tab:table4}. $a_k(\psi)=\exp(-\beta(\psi-\psi_k)^2)$, $\beta=0.05$. For Asn
$\psi_k=1.116$, $\delta_k=3836.80$}
\label{Ris16}
\end{figure}

Fig.~\ref{Ris16} demonstrates influence of the differences between $\mu_k\theta'_k$ and $\mu_k\sigma_k$ on the conductivities $\mu_k\theta'_k a_k$ and $\mu_k\sigma_k a_k$ when the concentration $a_k$, for example, has form $a_k=e^{-\beta(\psi-\psi_k)^2}$.

Note that directly in isoelectric point values of $\mu_k\theta'_k$ and $\mu_k\sigma_k$ are coincided
\begin{equation}\label{ZhSeq-C4}
\mu_k\theta'_k(\psi_k=\mu_k\sigma_k(\psi_k).
\end{equation}

Thus, in the steady state IEF process the concentration $a_k$ gives contribution to the conductivity of mixture only in the vicinity of the isoelectric point $\psi_k$ for two reasons. First, outside the vicinity of the isoelectric point there is no concentration of substances and, secondly, the effective conductivity of the substance outside the  isoelectric point vicinity is a very small. The most significant differences between the two forms of Ohm's law, of course, will be in the case of the non-stationary IEF process.

\begin{figure}[H]
\centering
\includegraphics[scale=0.8]{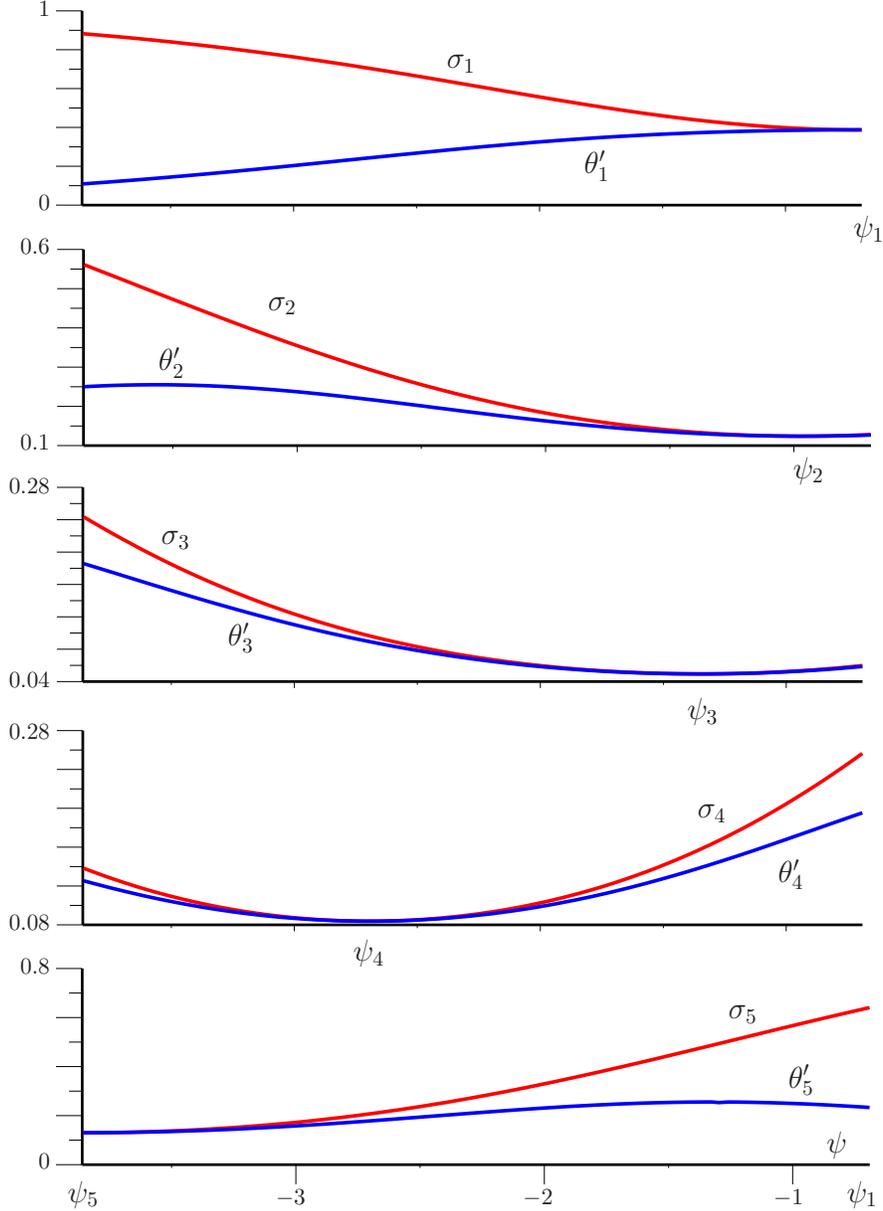}\\  [-4mm]
\caption{The differences between $\sigma_k$ and $\theta'_k$ for five-component
mixture. See table~\ref{tab:table4}}
\label{Ris17}
\end{figure}

In conclusion of this section we demonstrate (Fig.~\ref{Ris17}) differences  between $\mu_k\theta'_k$ and $\mu_k\sigma_k$ for five-component mixture (see
Tab.~\ref{tab:table4} and also \cite{Part2}) at $ \psi_1 \geqslant \psi  \geqslant \psi_5$.

\section{Stabilization of the electric current}\label{App:ZhS-4.4}

The boundary condition (see (\ref{ZhSeq-D5}))
\begin{equation*}
\varphi\left|_{x=0}\right.=\varphi_0 \quad \varphi\left|_{x=1}\right.=0
\end{equation*}
correspond to constant voltage IEF regime. In this case the current flux density $j$ depends on time: $j=j(t)$.

To realize the constant current flux density IEF regime one can change the value $\varphi_0$ calculating on each time step $j(t)$  and using formula
\begin{equation*}
\varphi_0(t+\tau)=\varphi_0(t)\frac{j_0}{j(t)}.
\end{equation*}

\begin{figure}[H]
\centering
\includegraphics[scale=0.85]{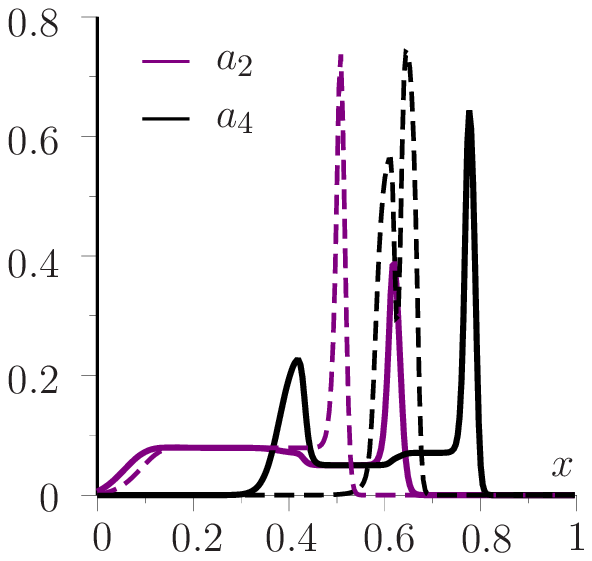}\
\includegraphics[scale=0.85]{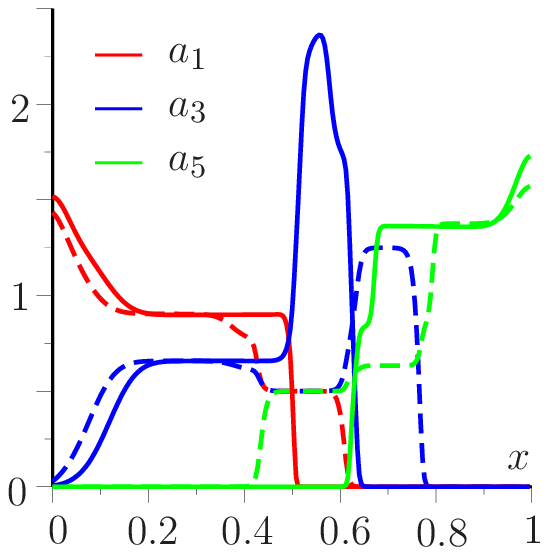}\
\includegraphics[scale=0.85]{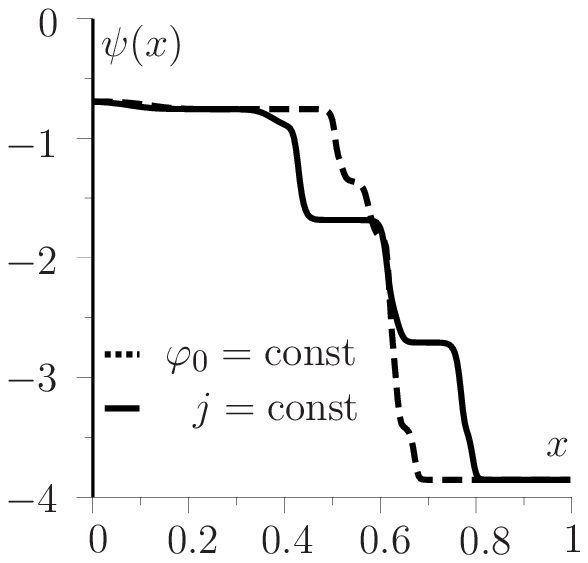}
\caption{Difference between constant voltage and current regimes at $\lambda=1500$, $t=110\tau$, $\tau=0.005$}
\label{Ris24}
\end{figure}

On Fig.~\ref{Ris24} the difference between these IEF regimes is shown.

\begin{figure}[H]
\centering
\includegraphics[scale=0.8]{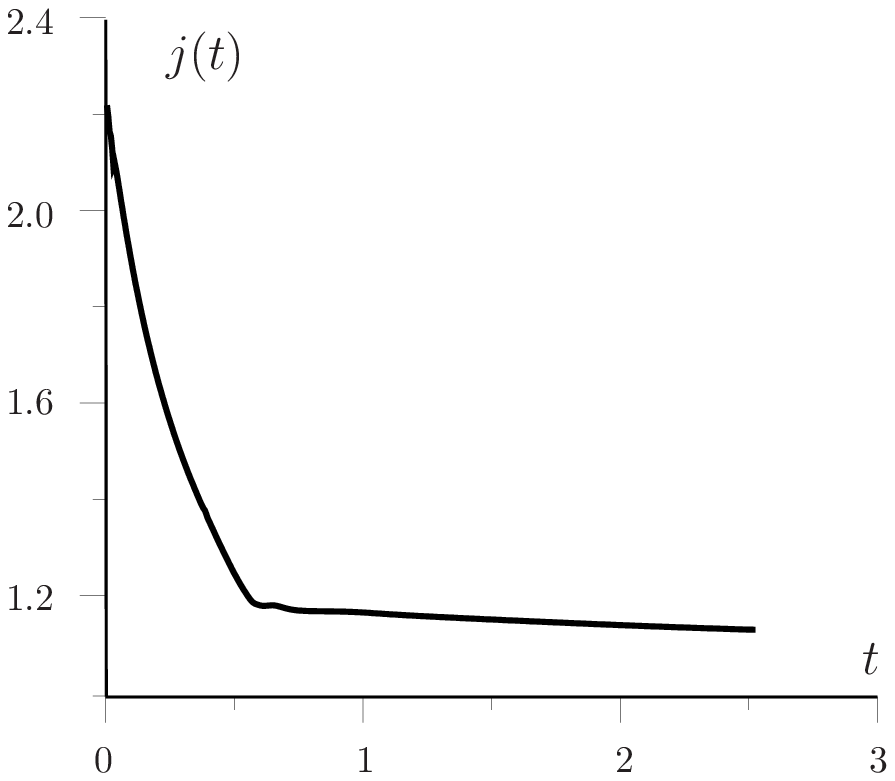}\
\includegraphics[scale=0.8]{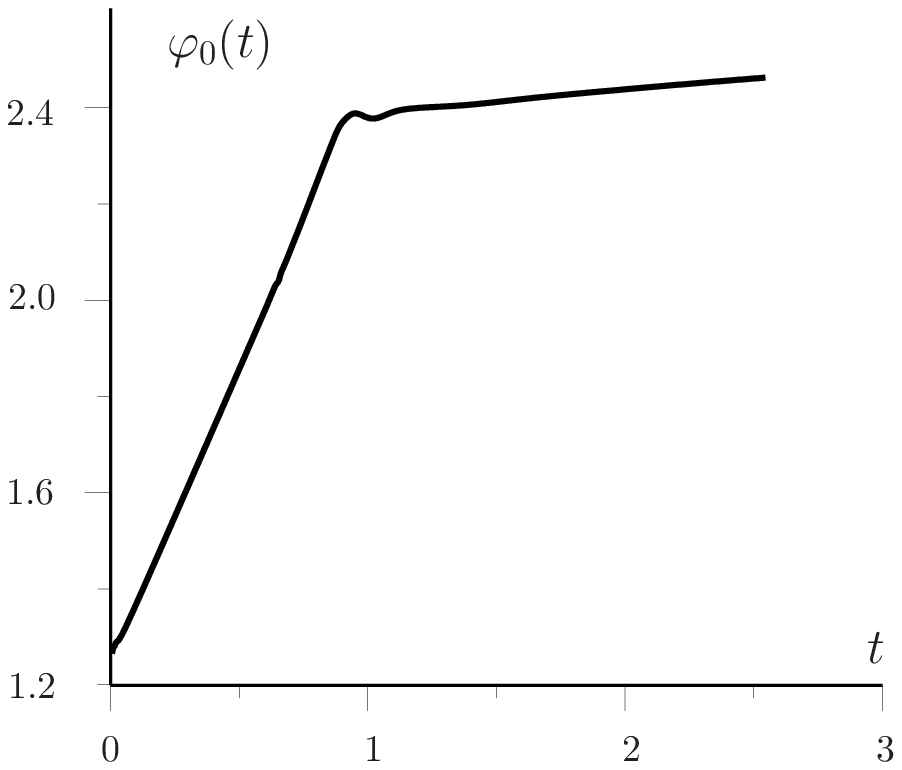}
\caption{Constant voltage ($U_0=2.808$, left) and current ($j=1.0$, right) regimes at $\lambda=1500$}
\label{Ris25}
\end{figure}

Analysis of the results shows that for the regime of constant current the fractionation of mixture occurs faster than for constant voltage regime.
The dependence of the current flux density and voltage on time for various regimes is shown on Fig.~\ref{Ris25}.

For constant current regime the time moment $t\approx 0.5$ corresponds to a significant rearrangement of the concentration profiles (see Figs.~\ref{Ris19}, \ref{Ris24}). For  constant voltage regime  the similar rearrangement at $t\approx 1$ is observed.

\section{Water ions influence on IEF}\label{App:ZhS-4.5}

The mathematical model (\ref{ZhSeq-D1})--(\ref{ZhSeq-D4}) does not take into account the influence of water ions on IEF. Contribution of ions $\textrm{H}^{+}$ and $\textrm{OH}^{-}$ in the molar charge and conductivity of mixture can play an important role to execute accurate calculations. The mathematical model, including the reaction of water dissociation, has the form (for more detail see equations (F27)--(F32) in \cite{Part4})
\begin{equation}\label{ZhSeq-D25}
\partial_t a_{k}+\operatorname{div} \boldsymbol{i}_k=0,\quad \boldsymbol{i}_k = -\varepsilon\mu_{k} \nabla a_k+\mu_{k} \theta_{k}(\psi)a_k \boldsymbol{E},
\quad k=1,\dots,n,
\end{equation}
\begin{equation}\label{ZhSeq-D26}
 \sum_{k=1}^{n}\theta_{k}(\psi)a_{k}+2K_w \sinh\psi=0,
\end{equation}
\begin{equation}\label{ZhSeq-D27}
 \operatorname{div} \boldsymbol{j}=0, \quad   \boldsymbol{E}=-\nabla \varphi,
\end{equation}
where
\begin{equation}\label{ZhSeq-D28}
\boldsymbol{j}=\sum_{k=1}^{n}
\left(
-\varepsilon \mu_k\nabla(\theta_{k}(\psi)a_k)+\mu_{k}\sigma_k(\psi)a_k \boldsymbol{E}
\right)+
\end{equation}
\begin{equation*}
+2k_w\mu_0\left(
-\varepsilon
\nabla(\sinh(\psi-\psi_0))+\cosh(\psi-\psi_0) \boldsymbol{E}
\right),
\end{equation*}
\begin{equation}\label{ZhSeq-D29}
\theta_k(\psi)=\frac{\sinh(\psi-\psi_{k})}{\cosh(\psi-\psi_{k})+\delta_{k}}=\frac{\varphi'_k(\psi)}{\varphi_k(\psi)},
\end{equation}
\begin{equation*}
\sigma_k(\psi)=\frac{\cosh(\psi-\psi_{k})}{\cosh(\psi-\psi_{k})+\delta_{k}}=\frac{\varphi''_k(\psi)}{\varphi_k(\psi)},
\end{equation*}
\begin{equation*}
\varphi_k(\psi)=\cosh(\psi-\psi_{k})+\delta_{k},
\end{equation*}
\begin{equation}\label{ZhSeq-D30}
   \psi_k=\frac12 \ln\frac{A_k B_k}{k_w^2}, \quad \delta_k=\frac12 \sqrt{\frac{B_k}{A_k}}, \quad
 \mu_{0}=\sqrt{\mu_{_H} \mu_{_{OH}}}, \quad
 \psi_0=\frac12\ln \frac{\mu_{_{OH}}}{\mu_{_H}},
\end{equation}
where  $\psi_i$ is the isoelectric point (electrophoretic mobility $\mu_i \theta_i$ is equal to zero at $\psi=\psi_i$, \emph{i.e.} $\mu_i \theta_i(\psi_i)=0$),
$\mu_0$ is the effective mobility of water ions,   $\mu_{_H}$, $\mu_{_{OH}}$ are the mobilities of hydrogen  ${\textrm H}^{+}$ and hydroxide $\textrm{OH}^{-}$ ions,
$\psi_0$ is the value of $\psi$ when water conductivity is minimal, $\delta_i>0$ is the dimensionless parameter, $\varphi_k(\psi)$ is some auxiliary function.
$K_w$ is autodissociation constant water.

The dimensionless values of the parameters $\mu_{_H}$, $\mu_{_{OH}}$, $K_w$,  $\mu_0$, and  $\psi_0$ are
\begin{equation}\label{ZhSeq-D31}
   \mu_{_H}=36.3, \quad   \mu_{_{OH}}=20.5,  \quad  \mu_{0}=27.2791, \quad
   \psi_0=-0.285696, \quad  K_w=10^{-6}.
\end{equation}

The calculation is convenient to hold for some hypothetical mixture. In this case, the solution has the additional symmetry properties and it is easier to interpret the results. We choose the following parameters close to the real five-component mixture
\begin{equation}\label{ZhSeq-D32}
   \mu_k=1.0, \quad   \delta_k=15,  \quad  \psi_{k}=6-k, \quad k=1,\dots,5, \quad U_0=15,
\end{equation}
\begin{equation*}
    M_1=0.125, \quad M_2=M_3=M_4=0.25, \quad M_5=0.125.
\end{equation*}

\begin{figure}[H]
\centering
\includegraphics[scale=0.8]{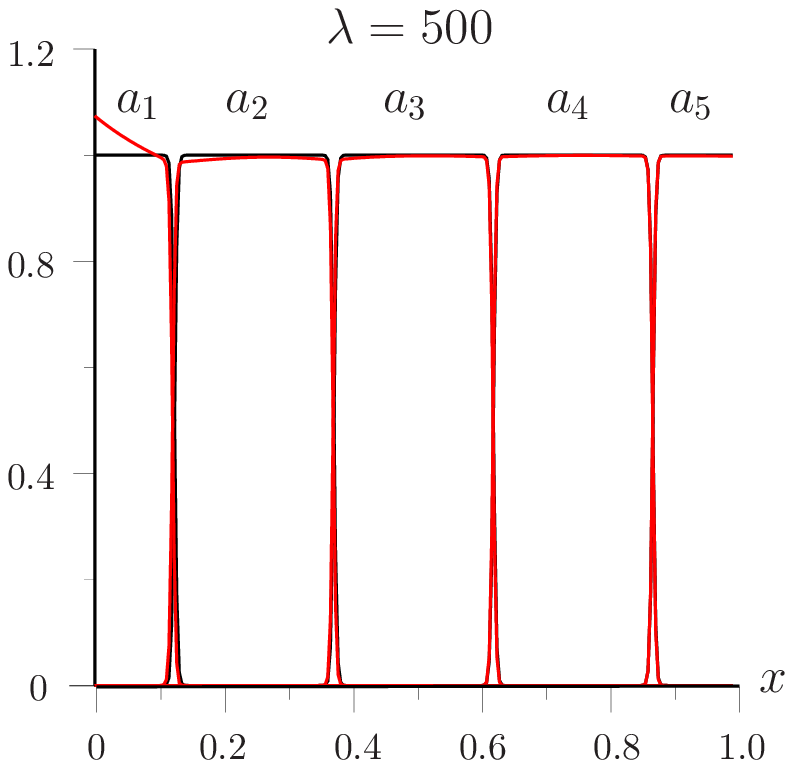}
\includegraphics[scale=0.8]{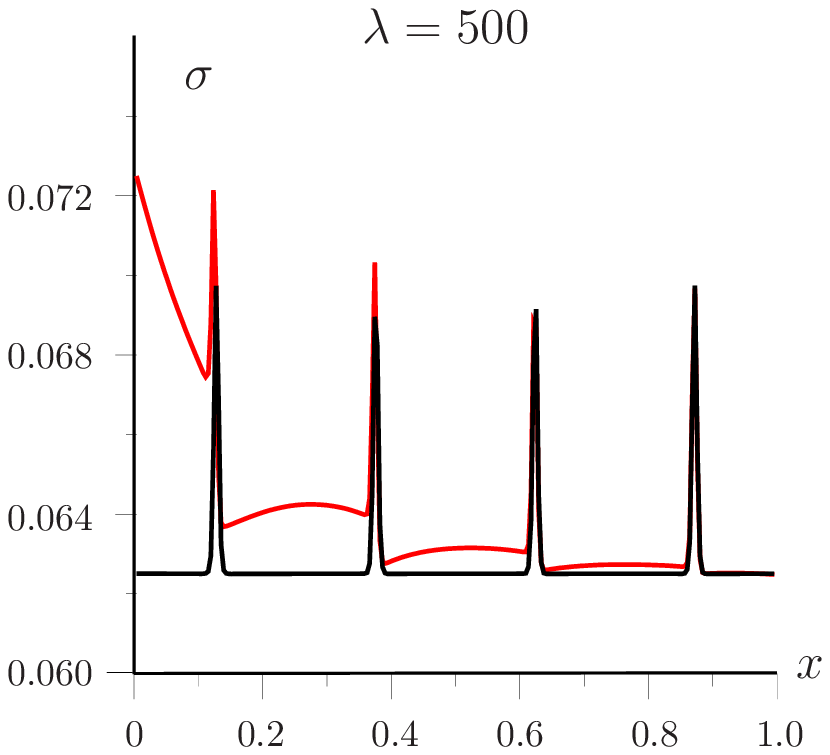}\\
\caption{The distributions of concentrations and conductivity.  $K_w \ne 0$, red line; $K_w = 0$, black line. The current constant regime ($j=1$) at $\lambda=500$,
$t=500\tau$, $\tau=0.005$, $L_*=2.54\,\textrm{cm}$}
\label{Ris26}
\end{figure}

On Fig.~\ref{Ris26} the result of calculation is shown.
The contribution of the water ions  is quite small due to the smallness parameter $K_w$. However, away from the values $\psi=0$ this contributions can be significant.

In particular, Fig.~\ref{Ris26} shows the distortion of concentration profiles as a result of water. The most obvious influence of water ions is observed on the conductivity of mixture, especially away from the values $\psi=0$ and for sufficiently large time (near a stationary state).  	
The distribution of acidity function $\psi(x)$ at $K_w \ne 0$ and $K_w = 0$  is almost the same.

\section{Influence of the differences between isoelectric and isoionic points on IEF}\label{App:ZhS-4.6}

To demonstrate influence of the differences between isoelectric ($\psi_k$) and isoionic ($\Psi_k$) points on IEF (see equations (F34)--(F36) in \cite{Part4}) we use the model (\ref{ZhSeq-D25})--(\ref{ZhSeq-D30})) changing $\theta_k(\psi)$ by $\Theta_k(\psi)$,
$\varepsilon \mu_k$ by $\mu_k \overline{\varepsilon}(\psi)$, and $\sigma_k(\psi)$ by $\overline{\sigma}_k(\psi)$ in equations (\ref{ZhSeq-D25}), (\ref{ZhSeq-D28}) (not in equation (\ref{ZhSeq-D26})!)
\begin{equation}\label{ZhSeq-D33}
\partial_t a_{k}+\operatorname{div} \boldsymbol{i}_k=0,\quad \boldsymbol{i}_k = -\mu_{k} \nabla(\overline{\varepsilon}(\psi) a_k)+\mu_{k} \Theta_{k}(\psi)a_k \boldsymbol{E},
\quad k=1,\dots,n,
\end{equation}
\begin{equation}\label{ZhSeq-D34}
 \sum_{k=1}^{n}\theta_{k}(\psi)a_{k}+2K_w \sinh\psi=0,
\end{equation}
\begin{equation}\label{ZhSeq-D35}
 \operatorname{div} \boldsymbol{j}=0, \quad   \boldsymbol{E}=-\nabla \varphi,
\end{equation}
where
\begin{equation}\label{ZhSeq-D36}
\boldsymbol{j}=\sum_{k=1}^{n}
\left(
-\varepsilon \mu_k\nabla(\Theta_{k}(\psi)a_k)+\mu_{k}\overline{\sigma}_k(\psi)a_k \boldsymbol{E}
\right)+
\end{equation}
\begin{equation*}
+2k_w\mu_0\left(
-\varepsilon
\nabla(\sinh(\psi-\psi_0))+\cosh(\psi-\psi_0) \boldsymbol{E}
\right),
\end{equation*}
\begin{equation}\label{ZhSeq-D37}
\theta_k(\psi)=\frac{\sinh(\psi-\psi_{k})}{\cosh(\psi-\psi_{k})+\delta_{k}}, \quad
\Theta_k(\psi)=\frac{\sinh(\psi-\Psi_{k})}{\cosh(\psi-\psi_{k})+\delta_{k}},
\end{equation}
\begin{equation*}
\overline{\sigma}_k(\psi)=\frac{\cosh(\psi-\Psi_{k})}{\cosh(\psi-\psi_{k})+\delta_{k}}, \quad
\mu_k\overline{\varepsilon}(\psi)=
\varepsilon\mu_k\Theta_k(\psi)-\varepsilon\mu_k^0\theta_k(\psi)+\varepsilon\mu_k^0.
\end{equation*}
Here, $\varepsilon\mu_k^0$ is the diffusive coefficient of the `neutral' ion $a_k^0$ (see equation (F5) in \cite{Part4}).

We consider again hypothetical mixture (\ref{ZhSeq-D32}) and choose
\begin{equation*}
   \Psi_1-\psi_1=-0.02, \quad    \Psi_2-\psi_2=0.01, \quad    \Psi_3-\psi_3=0.02, \quad    \Psi_4-\psi_4=0.01, \quad    \Psi_5-\psi_5=-0.01.
\end{equation*}

\begin{figure}[H]
\centering
\includegraphics[scale=0.8]{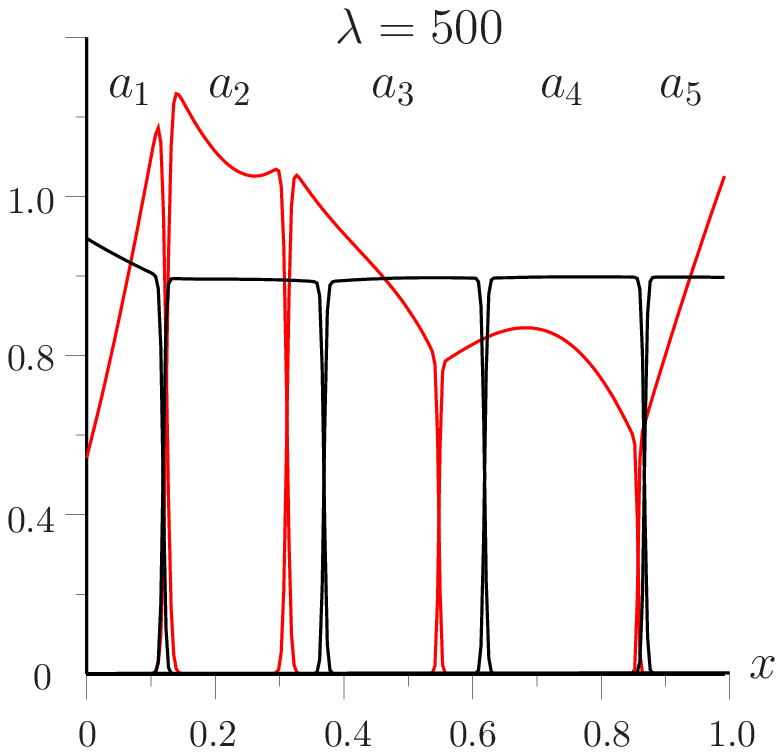}
\includegraphics[scale=0.8]{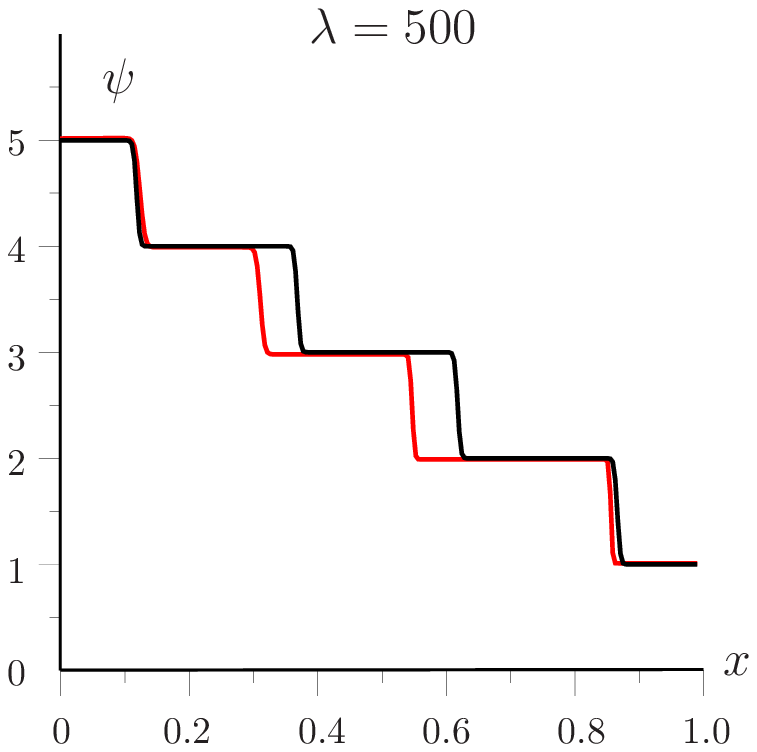}\\
\caption{The distributions of concentrations $a_k(x)$ and function $\psi(x)$ and conductivity at $K_w \ne 0$. The black line corresponds to the model (\ref{ZhSeq-D25})--(\ref{ZhSeq-D30}) and the red line corresponds to the model (\ref{ZhSeq-D33})--(\ref{ZhSeq-D37}). The current constant regime ($j=1$) at $\lambda=500$,
$t=500\tau$, $\tau=0.005$, $L_*=2.54\,\textrm{cm}$}
\label{Ris27}
\end{figure}

On Fig.~\ref{Ris27} the results of calculation are presented.
Although the differences $(\psi_k-\Psi_k)$ are small the profiles of concentrations vary appreciably. $\textrm{pH}$-gradient remains
piecewise constant function but moves in space and sizes of the sloping plot become different.  Such changes are especially visible at sufficiently large time moments, when the stationary state is almost reached.

The observed effect can be explained quite simply. In the stationary state the each concentration tries to be localized in a separate region. The mobility of the substance in this region tends to zero. However, due to the differences between $\psi_k$ and $\Psi_k$ molar charge of substances in region is non-zero. In this case, the distribution of the concentration is mainly determined by the electroneutrality condition. Note that the numerical simulations show that for large differences $(\psi_k-\Psi_k)$ the achievement of the stationary state is impossible.

\section{Implicit time discretization}\label{App:ZhS-4.7}

Described in section Sec.~\ref{App:ZhS-4.1} semi-implicit time discretization (\ref{ZhSeq-D8})--(\ref{ZhSeq-D11}) does not work  when the functions $\psi(x,t)$ and $\varphi(x,t)$ change according to the time quickly enough. The point is that the electroneutrality equation is solved  with poor accuracy. For accurate calculations, we should use the implicit scheme (compare equations (\ref{ZhSeq-D9})--(\ref{ZhSeq-D11}) with (\ref{ZhSeq-D39})--(\ref{ZhSeq-D41}) respectively)
\begin{equation}\label{ZhSeq-D38}
\frac{a_k^{(m+1)}-a_k^{(m)}}{\tau}+\partial_x i_k^{(m+1)}=0,
\end{equation}
\begin{equation}\label{ZhSeq-D39}
i_k^{(m+1)} = -\varepsilon\mu_{k} \partial_x a_k^{(m+1)}-\mu_{k} \theta_{k}(\psi^{(m+1)})a_k^{(m+1)} \partial_x\varphi^{(m+1)},
\end{equation}
\begin{equation}\label{ZhSeq-D40}
   \sum_{k=1}^{n}\theta_{k}(\psi^{(m+1)})a_{k}^{(m+1)}=0,
\end{equation}
\begin{equation}\label{ZhSeq-D41}
\partial_x \sum_{k=1}^{n}
\left(
\varepsilon \mu_k \partial_x(\theta_{k}(\psi^{(m+1)})a_k^{(m+1)})+\mu_{k}\sigma_k(\psi^{(m+1)})a_k^{(m+1)} \partial_x\varphi^{(m+1)}
\right)=0.
\end{equation}
In this case we should solve nonlinear equations (\ref{ZhSeq-D38})--(\ref{ZhSeq-D41}) instead of linear equations (\ref{ZhSeq-D8})--(\ref{ZhSeq-D11}). The computational practice showed that  it is possible to apply the simplest iterative process: instead solving the equations (\ref{ZhSeq-D38})--(\ref{ZhSeq-D41}) it is enough to solve at each time step cyclically the equations (\ref{ZhSeq-D8})--(\ref{ZhSeq-D11}) either convergence or limited to a few iterations. We confirm this statement showing the calculation results of the Kohlrausch's function  (see  equations (F39) and (F41) in \cite{Part4})
\begin{equation}\label{ZhSeq-D42}
R(x,t)= \sum_{k=1}^{n}\frac{a_k(x,t)}{\mu_k}.
\end{equation}
This function is particularly sensitive to the satisfaction of the  electroneutrality equations.

\begin{figure}[H]
\centering
\includegraphics[scale=0.8]{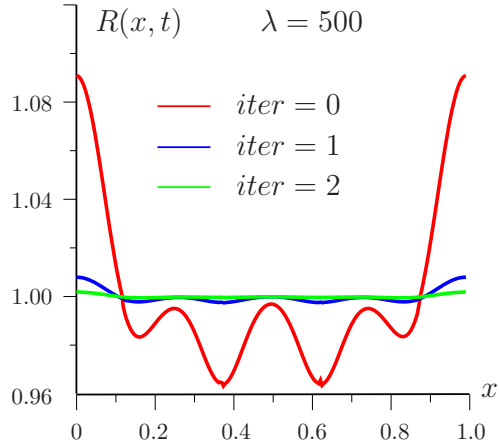}
\caption{The distributions of the Kohlrausch's function $R(x)$ at $K_w = 0$. Number iteration $iter=0,1,2$. The current constant regime ($j=1$) at $\lambda=500$,
$t=300\tau$, $\tau=0.005$, $L_*=2.54\,\textrm{cm}$}
\label{Ris28}
\end{figure}

On Fig.~\ref{Ris28} the Kohlrausch's function is demonstrated for the hypothetical mixture (\ref{ZhSeq-D32}).
Theoretical value of this function is $R(x,t)=1$. The $iter=0$ corresponds to semi-implicit time discretization. At  $iter=7$ we have $|R(x,t)-1|<0.0003$.

\section{Conclusion}\label{ZhS-13}

The results presented in this paper demonstrate that in the study of electrophoresis process we must use a more complete model. Neglect of those or other effects inevitably leads to incorrect results. The result of the use of simplified models may be inadequate compliance of the results of experiments. It is equally important to use the numerical schemes of high accuracy for calculations. Note that the presented model is still not have the required generality. In particular, this model does not take into account such important effects as the Wien's effects and the influence of ionic force of the solution on the mobility component of the mixture. An important result of the article is also the confirmation of the effectiveness of the use of the finite element method for solving the problem. Application of this method will allow to directly solve two-dimensional and three-dimensional problems for electrophoretic chambers with complex geometrical configuration. This is especially important in connection with the development of technologies Lab-on-Chip.

\begin{acknowledgments}
This research is partially supported by Russian Foundation for Basic Research (grants 10-05-00646 and 10-01-00452),
Ministry of Education and Science of the Russian Federation
(programme `Development of the research potential of the high school', contracts  14.A18.21.0873,  8832 and grant 1.5139.2011).

\end{acknowledgments}







\setlength{\bibsep}{4.0pt}

\end{document}